\documentclass[12pt,preprint]{aastex}
\usepackage{amsmath}

\begin{document}

\title{Modeling 
UX~Ursae~Majoris\footnotemark[1]~: An abundance of challenges.}
\footnotetext[1]
{Based on observations made with the NASA/ESA Hubble Space Telescope, obtained at the
Space Telescope Science Institute, which is operated by the Association of Universities 
for Research in Astronomy, Inc. under NASA contract NAS5-26555, and the NASA-CNES-CSA
{\it Far Ultraviolet Explorer}, which is operated for NASA by the Johns Hopkins University
under NASA contract NAS5-32985.}

%% Use \author, \affil, and the \and command to format
%% author and affiliation information.
%% Note that \email has replaced the old \authoremail command
%% from AASTeX v4.0. You can use \email to mark an email address
%% anywhere in the paper, not just in the front matter.
%% As in the title, you can use \\ to force line breaks.

\author{Albert P. Linnell$^2$, Patrick Godon$^3$, Ivan Hubeny$^4$, Edward M. Sion$^5$,
Paula Szkody$^6$}

\affil{$^2$Department of Astronomy, University of Washington, Box 351580, Seattle,
WA 98195-1580\\
$^3$Department of Astronomy and Astrophysics, Villanova University,
Villanova, PA 19085\\
visiting at the Space Telescope Institute, Baltimore, MD.\\
$^4$Steward Observatory and Department of Astronomy,
University of Arizona, Tucson, AZ 85721\\
$^5$Department of Astronomy and Astrophysics, Villanova University,
Villanova, PA 19085\\
$^6$Department of Astronomy, University of Washington, Box 351580, Seattle,
WA 98195-1580\\
}

\email{$^2$linnell@astro.washington.edu\\
$^3$godon@stsci.edu\\
$^4$hubeny@as.arizona.edu\\
$^5$edward.sion@villanova.edu\\
$^6$szkody@astro.washington.edu\\
}

\begin{abstract}
 We present a system model for optical and far UV spectra of the nova-like variable 
 UX~UMa	involving a white dwarf, secondary star, gas stream, hot spot and accretion 
 disk using our code BINSYN and based on an initially adopted system distance.
 Calculated SED intensity data successfully fit successive
 tomographically-extracted annuli longward of the Balmer limit but require
 a postulated `iron curtain' shortward of the Balmer limit that is applied to the
 annulus section closest to the secondary star, while postulated
 recombination emission fills in the model SED shortward of the Balmer limit and
 is applied to the annulus section more remote from the secondary star.
 The same model fits $UBV$ 1954 light curves by Walker and Herbig. Fits to
 $HST$ $FOS$ spectra are approximate but require assumed time-variable
 changes in the SED. Comparable effects, possibly involving variable absorption, afflict $FUSE$
 spectra. Fits to $IUE$ spectra by the model show time-dependent residuals
 that indicate changes in the accretion disk temperature profile, possibly indicative of a
 slightly variable $\dot{M}$ from the secondary star. Using model-based component
 light contributions and the improvement on the Bailey relation by Knigge we determine the system distance
 and mass transfer rate.

\end{abstract}

%% Keywords should appear after the \end{abstract} command. The uncommented
%% example has been keyed in ApJ style. See the instructions to authors
%% for the journal to which you are submitting your paper to determine
%% what keyword punctuation is appropriate.

%% From the front matter, we move on to the body of the paper.
%% In the first two sections, notice the use of the natbib \citep
%% and \citet commands to identify citations.  The citations are
%% tied to the reference list via symbolic KEYs. The KEY corresponds
%% to the KEY in the \bibitem in the reference list below. We have
%% chosen the first three characters of the first author's name plus
%% the last two numeral of the year of publication as our KEY for
%% each reference.

\section{Introduction}
\label{s-intro}

Cataclysmic variables (CVs) are
semi-detached binary stars in 
which a late-type typically main sequence star loses mass onto a white dwarf (WD) 
via Roche lobe overflow and  
accretion proceeds through a viscous disk. In the nova-like (NL) subclass, of which UX UMa 
is the prototype, the mass 
transfer rate from the secondary star 
is large enough that dwarf nova outbursts do not occur. 
(See \citet{warner95} for a history of UX UMa studies and a thorough review of CV types and properties.)
Important early studies of UX UMa by \citet{wh1954,jph1954} and \citet{kw1963} established the
system photometric properties, while \citet{nr1974} discovered coherent oscillations.
$IUE$ spectra identified UX UMa as a UV emission line variable \citep{holm1982,king1983}.
X-ray observations \citep{becker1981,wood1995} detected a soft source, further refined \citep{pratt2004} to
a noneclipsed soft component and a deeply eclipsed hard component. An accretion disk wind was
discovered by \citet{mason1995} and modeled by \citet{knigge1997}.

\citet{knigge1998a} had mixed success in modeling the spectral energy distribution (SED) of UX UMa
with a combination of stellar atmosphere synthetic spectra. A similar approach by \citet{froning2003}
(hereafter FR2003)
in a fit to the $FUSE$ spectrum produced an estimated $\dot{M}=1.6{\times}10^{-9}~M_{\odot}~{\rm yr}^{-1}$.
However, a mass transfer rate of
$\dot{M}=1.6{\times}10^{-9}~M_{\odot}~{\rm yr}^{-1}$ is small enough that the system would be
unstable against outburst, contrary to observation.
Neither of these two studies included a SED contribution from the WD;
on the other hand, tomographic reconstruction of the UX UMa disk \citep{ru1992} is consistent with
a standard model accretion disk with $\dot{M}=5{\times}10^{-9}~M_{\odot}~{\rm yr}^{-1}$.
A comparable study by \citet{bap1995} hereafter BA1995, with a different estimated distance to the system, finds
$\dot{M}=1{\times}10^{-8}~M_{\odot}~{\rm yr}^{-1}$.
\citet{frank1981} modeled IR and optical light curves and determined system parameters.	
\citet{smak1994a} modeled optical region light curves and derived bright spot parameters. 
\citet{mason1997} argue
that a more complicated structure than a bright spot on an 
otherwise unmodified accretion disk rim is necessary. 
BA1995 derived a WD $T_{\rm eff}$ of 52,000K to 70,000K, but FR2003 points out that a
$T_{\rm eff}$ that large is inconsistent with FUV spectra.
Neither of the tomographic reconstruction studies has been the basis of an effort to model
the spectra.
These continuing inconsistencies are the motivation to attempt to derive a single model that represents
a substantial part of the disparite observational data.    

\section{Observational Data: Spectroscopic and Photometric Observations}

Table~1 lists the spectroscopic observations used in this study. 
We obtained the $IUE$ spectra from the $IUE$ archive and processed them with the standard
IUEDAC package.
The $HST/FOS$ observations are from \citet{knigge1998a}, further analyzed in
\citet{bap1998}, hereafter BA1998, and described in that publication: 
the G160L data sets each include 691 spectra and
the PRISM data sets each include 870 spectra; these time-resolved spectra have a resolution of 5.3 s.
We obtained the time-integrated spectrum of each of the G160L data sets from the MAST archive for
our analysis and designate this data set DS1.
We used
spectroscopic data from the \citet{ru1994} paper, 
consisting of optical spectra obtained in June 1992; we designate this data set DS2.
For that data set we used the Dexter facility of the
Astrophysics Data Service (ADS) to digitize the spectrally-resolved disk spectra for regions A-F in 
their Figures~5a-f. 

In agreement with \citet{ru1992}, we adopt an interstellar reddening of $E(B-V)=0.0$.
\citet{bruch1994} list $E(B-V)=0.02$. This small value is uncertain because of the variability of the system
light curve discussed below and we prefer to leave the reddening at 0.0 for our analysis.
Unless stated otherwise, changing to $E(B-V)=0.02$ has a negligible effect on our results.
The $FUSE$ spectra, from the MAST archive (and used by FR2003), were processed with CalFUSE v3.1. 
The dataset includes 
multiple exposures; we
have selected two for this study.
The $HST/FOS$ spectral lines were identified and studied by \citet{knigge1998a} and BA1998 while
the $FUSE$ spectral lines were identified and studied by FR2003; we do not repeat those identifications.

\section{Initial System Parameters}

\citet{ru1992} list several determinations of the distance to UX UMa and adopt a compromise value
of 250pc. Additional determinations are
\citet{smak1994a} ($328{\pm}12$pc), \citet{frank1981} ($340{\pm}110$pc), and BA1995 ($345{\pm}34$pc). 
FR2003 challenge 
the BA1995
system parameters, and our analysis agrees with some of the FR2003 points. 
(We have concerns about the BA1995 determination; see \S6.)
We choose to adopt the distance of 250pc for our working model, initially as an illustrative test and
subject to subsequent consistency verification.	The distance uncertainty remains a major obstacle and
there is a degeneracy between the
system distance and the adopted $\dot{M}$. 

\citet{ru1994} determined an approximate spectral type of M0 for the secondary star in UX UMa and a
corresponding main sequence mass of ${\sim}0.5M_{\odot}$. This determination, together with the
orbital period, is consistent with the \citet{smak1994a} system parameters of 
$M_{\rm wd}=0.70{\pm}0.2M_{\odot}$, $M_{\rm s}=0.49M_{\odot}$, and $q=0.70$. The period-secondary mass relation of
\citet[eq. 2.100]{warner95} gives a secondary mass of $M_{\rm s}=0.48M_{\odot}$, while the \citet{knigge2006}
period-secondary mass relation finds $M_{\rm s}=0.44M_{\odot}$.
A radial velocity study by \citet{sh1984} finds that 
$M_{\rm s}{\la}0.5M_{\odot}$ and the mass ratio of the system is greater than ${\sim}0.8$.
\citet{frank1981} find $0.2<M_{\rm s}/M_{\odot}<0.5$ and $0.1<M_{\rm wd}/M_{\odot}<0.5$.
They also find $\dot{M}=8{\times}10^{-9}~M_{\odot}~{\rm yr}^{-1}$ (see below). It is worth mentioning
that, at an orbital period of $4.7^h$, UX UMa lies in a region of parameter space where nuclear
evolution of secondaries becomes significant, adding a measure of uncertainty to standard relationships.
 
 The BA1995 $M_{\rm wd}=0.47M_{\odot}$ depends on the radial velocity value
 $K_{\rm wd}=160~{\rm km}~{\rm s}^{-1}$ \citep{sh1984} determined from the emission wings of H$\alpha$.
 \citet{sch1983} found a range of $K_{\rm wd}$ values for different spectral features.	FR2003 used 
 a cross-correlation procedure
 to determine a $K_{\rm wd}$ of $70~{\rm km}~{\rm s}^{-1}$, while absorption reversals in emission lines determined
 $K_{\rm wd}$ values from $140~{\rm km}~{\rm s}^{-1}$ for S~IV and S~III to $200~{\rm km}~{\rm s}^{-1}$ for C~III.
 Thus there is some uncertainty in the $M_{\rm wd}$ determination; a more secure value of $M_{\rm wd}$ depends on
 an improved radial velocity curve attributable to the WD. 
 Note that the value $q=1$ is at the limit of stability
 against dynamical mass transfer which occurs for $q>1$. 
We provisionally adopt the BA1995 model, $M_{\rm wd}=M_{\rm s}=0.47M_{\odot}$. This model	closely
agrees with the model adopted by \citet[Table~4]{ru1992}, $M_{\rm wd}=M_{\rm s}=0.45M_{\odot}$.

From \citet{p2000}, the zero temperature radius of a $0.47M_{\odot}$ WD is 
$R_{\rm wd,0}=0.01375R_{\odot}=9.57{\times}10^8{\rm cm}$ and, 
from their Fig.4a, we find that the radius of a 20,000K,
$0.47M_{\odot}$ WD, used later in our discussion,
would be $1.15{\times}10^9{\rm cm}$. 
BA1995 argue that, if 1/2 of the WD is visible,
it has a $T_{\rm eff}=70,000$K. This temperature is off the Figure~4a plot of \citet{p2000}, but we estimate 
a radius of $3.5{\times}10^9{\rm cm}$. A 70,000K WD with this radius (and visible to the observer)
would completely dominate 
the system flux and would lead to a seriously discrepant synthetic spectrum (see below).	
(Also see the discussion by FR2003.)

Table~2 lists the system parameters	used in this paper. Except for $i$ the parameters are adopted as discussed
herein and so have no errors attached.
The Roche potential for the WD produces a radius appropriate to a $0.47M_{\odot}$, 20,000K WD.
The secondary star Roche potential causes it to fill its Roche lobe.
\citet{ru1992} find $i=71{\arcdeg}$ for their $q=1.0$ model. They investigate the sensitivity of this
result to variation of $q$ and $i$, and find their result is not very sensitive provided the eclipse
width remains unchanged. In our light curve simulations (\S4.1) we found a slight
improvement in the fit to eclipse depth with $i=70.2{\arcdeg}$, but a detectably poorer fit to
eclipse width with $i=70.0{\arcdeg}$. We adopt $i=70.2{\arcdeg}$.

Using a maximum-entropy eclipse mapping algorithm (the MEM method), \citet{ru1992} determine a 
temperature profile for
the UX UMa accretion disk (their Figure~4b). The procedure uses four optical-wavelength eclipse light
curves that constitute the observational data. A two-dimensional array of surface elements (pixels)
covers the model accretion disk which is projected on the sky at inclination $i$. A default intensity
is assigned to each pixel at each of the observed wavelengths and an assumed distance together with
calibrated absolute magnitudes for the observational data produces intensities in physical units. 
The algorithm then iterates to vary
the individual intensities until the entire calculated array of pixel intensities has maximum entropy, thereby
fitting the observed light curves with calculated light curves. A black body fit to individual pixel
intensities at the four observed wavelengths determines $T_{\rm eff}$, pixel by pixel. The temperature
profile then is a plot of the pixel $T_{\rm eff}$ values as function of the pixel distance from the
center of the WD.

We used the ADS Dexter facility to digitize the
points in their plot. 
We also used the Dexter facility to digitize the BA1998 `back' points (discussed in \S4.) from their
Figure~7 for November 1994.
Figure~1 compares the temperature profile in our model with the points
from Rutten et al. (crosses) and Baptista et al. (triangles). Note that the plotted points do not
represent directly observed quantities; rather, they are derived quantities from the maximum
entropy maps.
Based on the Figure~1 plot, our model adopts a standard model \citep{fkr92}
$\dot{M}=5.0{\times}10^{-9}M_{\odot}{\rm yr}^{-1}$ temperature profile as an initial approximation.	
By eye estimate, the temperature profile for our model is a fairly good fit to the plotted data points.
The data points depart from	a standard model near the L1 point where the bright spot and tidal heating
exert an effect (discussed below). 

There are no data points at extremely small radii; we find 
ambiguous evidence for obscuration of the WD and/or for a hot boundary layer, raising doubts about the
justification of modeling the theoretical temperature downturn at very small radii. The theoretical
downturn assumes a slowly-rotating central star \citep{fkr92}; with faster rotation the downturn
becomes smaller and we have no information on the rotation rate of the WD.
Our Figure~1 profile shows only the start of a temperature drop near the WD and so fails to fit the 
theoretical $\dot{M}$ profiles at small radii.	
Our model calculates only one annulus with
radius less than the radius of the temperature maximum at $R=1.36R_{\rm wd}$. 

Our program uses the \citet[Equation~2.61]{warner95} relation for the tidal cutoff radius and for $q=1.0$ this leads
to $R_{\rm cutoff}=0.6R_{\rm L1}$. \citet{ru1992} and BA1998 both tabulate data
points to $R=0.7R_{\rm L1}$. This difference explains the failure of our model to extend to the tabular
terminus of the Rutten et al. data in Figure~1. We subsequently (\S4.1) model light curves with the aid of a bright spot
on the accretion disk rim. \citet[Table~7]{ru1992} determine $R_{\rm spot}/R_{\rm L1}=0.57{\pm}0.05$, in
agreement with the rim location in our model. If our model allowed the accretion disk to extend to 
$R_{\rm cutoff}=0.7R_{\rm L1}$
the $T_{\rm eff}$ of the outermost part of the accretion disk would be below the temperature limit for stability 
\citep{osa96}. We do not believe the tidal cutoff sets a sharp boundary beyond which no emitting gas can be found
but for formal internal consistency	we adopt $R_{\rm cutoff}=0.6R_{\rm L1}$.

\subsection{Calculation of synthetic spectra}

 Our model uses the BINSYN suite (\citet{linnell1996,linnell2008} and references therein).
 This model includes contributions from the accretion disk, the WD, the secondary star, and the
 accretion disk rim.
 \citet{smak2002} shows that the structure of the outer parts of accretion disks is an important consideration.
 Smak shows that allowance for tidal heating produces a rim $T_{\rm eff}$ comparable to that of the
 outermost annulus. On a finer scale, which Smak does not discuss in detail, the part of the rim close to the
 orbital plane has a higher $T_{\rm eff}$ than at a higher $z$ value. We set the rim $T_{\rm eff}$ equal to
 that of the outermost annulus and make the rim isothermal. 
 The program requires a set of individually-calculated accretion disk annulus spectra which are based on
 annulus models calculated with program TLUSTY \citep{h1988,hubeny1990,hl1995,hh1998}.
 The annulus models do not include irradiation by the central star although this is a TLUSTY option; 
 we will find there is some
 uncertainty concerning the WD $T_{\rm eff}$.
 We emphasize that the TLUSTY option for accretion disk annuli (without irradiation) produces a standard 
 model radial temperature
 profile and includes explicit treatment of the mass transfer rate and the WD mass and radius. 
 It would be impractical to recalculate TLUSTY annulus model arrays for different WD radii depending on the radii
 appropriate to different assumed $T_{\rm eff}$ values.	We calculate a single array appropriate to
 a zero temperature WD and handle the effect of a non-zero temperature WD in the BINSYN model.
 Table~3 lists the annuli used in our model and gives some of the individual annulus properties.
 The first column lists the radius of the annulus; the second column lists the $T_{\rm eff}$; the third
 column lists the column mass to the central plane in ${\rm gm/}{\rm cm}^2$; the fourth column
 lists the temperature at the cental plane; the fifth column lists log~$g$ (cgs) at a Rosseland optical depth
 of approximately 0.6; the sixth column lists the vertical distance above the central plane of the
 optical depth 0.6 point, in cm; the final column lists the Rosseland optical depth of the central plane.
 The annulus model calculations through $r/r_{\rm wd,0}=12.0$ converged. The remaining models are grey model
 solutions and are subject to some scatter in the calculated $z_{\rm H}$.	
 The TLUSTY models have standard model \citep{fkr92} $T_{\rm eff}$ values.

 Program SYNSPEC \citep{hlj1994} uses the TLUSTY output to calculate individual annulus synthetic spectra.
 The spectra were calculated with a spectral resolution of 1\AA. An important point is that SYNSPEC
 produces both of two 
 different output formats. In one, the output is in Eddington flux units 
 (${\rm erg}~{\rm cm}^{-2}~{\rm s}^{-1}~{\rm \AA}^{-1}$). In the other the output is in intensities
 (${\rm erg}~{\rm cm}^{-2}~{\rm s}^{-1}~{\rm hz}^{-1}~{\rm sr}^{-1}$) for a specified array of directions
 relative to the outward normal. In our calculations we specified ten directions equally spaced in
 $cos({\gamma})$, where ${\gamma}$ is the angle between the chosen direction and the outward normal to the
 local photosphere. We will specifically require the intensity-based data for some of the subsequent discussion.

 BINSYN sets up a separate array of annuli to represent the accretion disk; in this study we used 45
 division circles to produce 44 annuli, with inner radius at the WD boundary and outer radius at the
 tidal cutoff boundary. This choice provides adequate radial resolution for accurate calculation of
 both synthetic spectra and light curves. One half of the WD is visible in this model.
 BINSYN uses the \citet[eq.~5.45]{fkr92} equation set to calculate 
 a standard model accretion disk, including the rim height.
 Table~4 lists the array properties. In calculating flux from annulus segments we use the $T_{\rm eff}$
 of the annulus inner edge. 
 The tabulation of accretion disk annuli from TLUSTY, Table~3,
 begins at the accretion disk temperature maximum since following the theoretical temperature drop to
 smaller radii would lead to double-value ambiguity (either side of the maximum) in the BINSYN interpolation.
 The algorithm for assigning radii in BINSYN places at least one division circle interior to the
 temperature maximum. The two initial entries with the same $T_{\rm eff}$ represent the inner and outer radii
 of the first annulus; thereafter the table lists the outer radius.

 The $T_{\rm eff}$ values in Table~4 are appropriate for a 20,000K WD, with its radius determined
 as in \S3. The change in WD radius from a zero temperature WD produces a clearly detectable change in the
 tabular values of the BINSYN annuli $T_{\rm eff}$ values. Compare the $T_{\rm eff}$ values at
 $r/r_{\rm wd}=1.36$ in Table~3 and Table~4; the difference is due to the different values of $r_{\rm wd}$.
 BINSYN determines flux values for segments on a given
 annulus by using the Table~4 $T_{\rm eff}$ value to interpolate, temperature-wise, within Table~3,
 making use of the flux values corresponding to the Table~3 entries. Note that Table~3 is based on
 $r_{\rm wd,0}=9.57{\times}10^8~{\rm cm}$ while Table~4	is based on 
 $r_{\rm wd}({\rm T=20,000K})=1.15{\times}10^9~{\rm cm}$.
 
 By option, the Table~4 profile may be replaced by a different non-standard model profile. This option
 has been used in a number of instances (e.g., \citet{linnell2008}).

 \section{Comparison of model with DS1 and DS2}

 Our comparisons will use eclipse maps associated with the DS1 and DS2 data and these maps will require default
 intensities for the map pixels.
 Calculation of default intensities requires a source distance to determine the subtended solid angle.
 DS1 adopted a distance of 345pc; DS2, 250pc. Our working model adopts 250pc (\S3.), so
 to produce a common standard of comparison we divide the first data set intensities by $(345/250)^2=1.9$.
 Separate tests show that our results remain essentially unchanged if we adopt a distance of 345pc,
 normalize observed intensities to that distance, and adopt ${\dot{M}}=1.0{\times}10^{-8}M_{\odot}~{\rm yr}^{-1}$.

Production of a MEM eclipse map, which determines intensities pixel by pixel, permits segregation of pixels 
into isolated regions on the accretion disk.
Thus, DS1 divides the accretion disk into concentric rings, each ring further subdivided into a 'back', a 'front',
and a 'stream'.	Figure~2a illustrates the DS1 divisions. From \S2, the observational data include
both the PRISM and G160L as separate spectral sources. Consequently, it was necessary to introduce identifiers for
the source of the data used as well as the position of the ring subdivision.
  Thus, in DS1,
 identifier `prism.center' refers to the radial range 0.000$r_{\rm L1}$-0.075$r_{\rm L1}$, while the identifier
 `prism.back.20' refers to the azimuthal range $90{\arcdeg}$ to $270{\arcdeg}$ on the annulus
 extending from $0.175r_{\rm L1}$ to $0.225r_{\rm L1}$. Table~5 lists the subdivisions of the accretion disk
 and may be compared with Figure~2a. DS2 follows a similar but coarser scheme of geometric resolution; its 
 subdivisions also are listed
 in Table~5.
 
 The PRISM spectra were divided into 127 passbands and a separate MEM map was constructed for each passband.
 For each ring subdivision, concatenation of the 127 passband intensities constituted an extremely low resolution
 spectrum. It is that 'spectrum', for a given ring subdivision, that we fit with our model synthetic spectrum.
 Since the data being fit consist of intensities, it is necessary to use the corresponding intensity output from 
 SYNSPEC as discussed
 in \S3.1. 
 
 The G160L spectra were divided into 59 passbands and
 processed as with the PRISM data. DS2 used low-resolution optical wavelength spectra and followed a similar
 basic procedure.  
 Note that DS2 are corrected for extinction in the Earth's atmosphere
 while DS1 requires no such correction.
 
 Figure~2b illustrates the BINSYN model and the 44 annuli. Software keys isolate annuli subdivisions
 ('front', 'back','stream') for comparison with DS1. 
 Given the Table~5 radial ranges, in each comparison we select an annulus from Table~4 that is centered 
 on the DS1 or DS2 annulus in question.
 
 The \S3 system parameters and the $\dot{M}$ from Figure~1 set the model scale and the
 radiation properties of the individual photospheric segments in physical units. Production of a system 
 synthetic spectrum
 involves integration over the entire (projected) system model. Software keys restrict the contributing
 segments to corresponding DS1 or DS2 regions. The final step is division of the integrated model intensity
 by the contributing model area. There are no adjustable parameters to apply to the model in a comparison
 with the DS1 or DS2 data. We stress that the DS1 and DS2 data are not directly observed quantities but
 result from MEM maps which require an assumed distance for expression in physical units.

 Both data sets provide intensities for a central region that includes the WD; we have found that the
 most consistent overall choice for the WD $T_{\rm eff}$ is $\approx20,000$K. Figure~3 compares our model
 with DS2 region `A', Table~5.
 The error bars are an average from the 
 \citet{ru1994} Table~1. 
The reasonably close agreement between our model and
 the region 'A' data is gratifying. 
 Based on the error bars, the discrepancy shortward of
 5000\AA~probably is real (i.e., not a data artifact such as, e.g., incomplete correction for atmospheric
 extinction at a 2$\sigma$ level) but
 we remain
 cautious in interpreting the discrepancy because of the results for Figure~4a and Figure~4b, discussed
 below. The discrepancy could be explained by an opacity source not included in our model.

 Figure~4a compares a 40,000K WD model (upper synthetic spectrum), a 20,000K WD model (middle synthetic spectrum), 
 and a zero temperature model (lower synthetic spectrum) with the G160L 
 and prism data. The synthetic spectrum for the accretion disk contribution corresponds to BINSYN annuli
 1-6. Minor corrections, not discussed here, 
 would raise the 40,000K synthetic
 spectrum close to the peaks of the observed spectra.

 BA1998
 note that there was a $70\%$ change in $\dot{M}$ between August 1994 (G160L) and November 1994 (prism)
 in the sense that the disk intensities are larger for the prism data. 
 We empirically found that the
 G160L spectra accord well with the prism spectra if the G160L intensities are divided by 
 $1.2{\times}10^{-3}$ (and used in the plots) rather than 
 the $1.9{\times}10^{-3}$ factor used with the prism spectra.

 The 40,000K WD contributes substantially to the upper synthetic spectrum 
 while the inner accretion disk annuli are the major contributors to the 20,000K WD model; the data
 do not discriminate the latter model from a zero temperature WD model. 
 (We produced the system synthetic 
 spectrum for a 
 zero temperature WD
 by suppressing the WD contribution to the integrated intensity but left the total
 projected area unchanged.)
 Note that the
 peak intensity for the 20,000K WD model is at an ordinate value of 3.5, substantially discrepant from 
 the peak observed intensity.
 The Figure~4a data are in conflict with the Figure~3 data. The WD $T_{\rm eff}$ is one of the
 principal issues to be addressed and the Figure~4a data (plus the Figure~4b data: see below) constitute
 the only direct evidence for a 40,000K WD while the remaining data, discussed below, provide inconclusive
 (\S6.) support for a 20,000K WD.
 
 Figure~4b shows the continuation of 
 Figure~4a to longer wavelengths. Note the different spectral gradient at wavelengths longward of about 
 5500\AA~but possibly extending to shorter wavelengths. 
 BA1998 describe a dropoff in the prism spectrograph response in that region and beyond while 
 the DS2
 data set (Figure~3) agrees with our model; 
 the prism data appear to have a calibration issue (BA1998) that is insignificant at the short wavelength end
 but increases to longer wavelengths.

 The Figure~4a error bars are small and the DS1 fractional errors gradually
 increase as the measured intensity drops in successive figures; to avoid excessive clutter we 
 provide DS1 error bars only for
 Figure~4a and Figure~4j.
 Neither synthetic spectrum (the 20,000K case or the 40,000K case in Figure~4a) models the observed 
 deep absorption features of Si II ${\lambda}1300$;
 Si IV ${\lambda}1394,1403$; C IV ${\lambda}1548,1552$; He II ${\lambda}1640$; N V ${\lambda}1718$;
 Si III ${\lambda}1892$ and Mg II ${\lambda}2800$. It is noteworthy that the C IV ${\lambda}1548,1552$
 doublet is in emission; \citet{knigge1997} show that there is a wind
 associated with the emission feature and that there is an underlying slow-moving `chromosphere' that can produce
 narrow absorption reversals.
 The question arises whether a change in the physical conditions in the system, in the two-year interval
 between the Figure~3 data (obtained in 1992) and the Figure~4a data (obtained in 1994), could explain 
 the discrepancy between the two figures.
 There is evidence for a high temperature source located near where
 a transition layer would be expected. The evidence includes a hard X-ray source \citep{pratt2004},
 an eclipsed flickering source \citep{bruch2000}, a 29-second oscillation source that is eclipsed 
 \citep{knigge1998b}, and a source of excess radiation shortward of 965\AA~(see below). 
 Further discussion of the WD $T_{\rm eff}$	and the hot source is in \S6.

 Figure~4c shows the DS1 data for Table~5 radial designation 0.10.
 In this and subsequent plots the heavy grey line is the 'front' region, the light grey line is the
 'back' region and the light continuous line is the 'stream'. The heavy continuous line is the model
 synthetic spectrum. Note that the synthetic spectrum fits the DS1 data very well near 4000\AA. The
 'back' shows a higher intensity than the 'front' and the synthetic spectrum lies between them. At
 this annulus the 'stream' is in fair agreement with the 'front'.

 Baptista et al. discuss the lower flux levels in the 
 `front' annulus regions as compared with the `back'. They propose absorption due to a large
 number of blended lines of FeII similar to the `iron curtain', hereafter IC,
 invoked by \citet{hor1994}, hereafter HO1994. The disk $T_{\rm eff}$ at this radius is about 21,000K. 
 The IC calculated by HO1994 for
 OY Car (their Figure~8) produced absorption shortward of the Balmer discontinuity and 
 increased the depth of the Balmer jump. Thus, a similar mechanism is credible to explain
 the `front' discrepancy from the model synthetic spectrum.
 There is
 excess flux in the `back' Balmer continuum, and there is no Balmer jump.
 \citet{knigge1998a} propose H recombination emission from an accretion disk chromosphere (ADC) as a
 mechanism to fill in the Balmer discontinuity. Their ADC is at the base of a biconical wind and
 has a vertical height of order $10R_{\rm wd}$ to provide an emission measure adequate to fill
 in the Balmer jump. 
 Note the slight difference between the amount of absorption for the `stream'
 and the 'front`.  The ADC, the IC, and the biconical wind probably are part of a single structure.
 The IC absorbing material most likely is located above the outer part of the accretion disk and is
 seen projected on the inner annuli where most of the radiative flux is produced. 

 The problem with the prism data calibration is apparent. If it were possible to adjust the synthetic spectrum
 fit empirically, the wavelength at which dropoff starts could be moved to a longer wavelength from
 around 5000\AA. There are no adjustable parameters; it would otherwise make sense to truncate
 the prism spectrum at some appropriate wavelength but the fit changes from figure to figure so there is no
 good way to choose a truncation wavelength. These plots, including their agreement with the
 DS2 data (see below), serve as documentation of the prism
 calibration problem.

 Figure~4d compares the model with prism data for radial designation 0.15. 
 The dotted line is the data for DS2 region `B'. 
 Note the
 agreement of the DS2 profile with the model spectrum. As in Figure~4c, the model spectrum,
 shortward of the Balmer limit, is intermediate between the `front' and `back' and 
 is in excellent agreement with the `stream'. Ignoring the unmodeled emission lines, the `front' Balmer
 jump agrees with the synthetic spectrum. The two mechanisms of Figure~4c (IC and recombination emission)
 also may explain the
 residuals here.

 Figure~4e compares the model with prism data for radial designation 0.20.
 As in Figure~4d, shortward of the Balmer limit, the model spectrum is intermediate
 between the 'front' and 'back' but the 'stream' now lies closer to the 'back'.
 
 Figure~4f shows data for radial designation 0.25. 
 The dotted line is the DS2
 annulus `C' spectrum.
 In contrast to Figure~4d, the DS2 data indicates a higher intensity than the
 corresponding DS1 data, while the model spectrum lies half way between the two data plots.
 A reduction in the model $T_{\rm eff}$ from the 11,307K of Table~4 annulus \#~19 to 10,898K
 produces a close fit to the DS1 data longward of 4000\AA~and a fairly close fit shortward, 
 consistent with a local
 departure from the standard model; a corresponding but smaller increase of the model $T_{\rm eff}$
 produces a close fit to the DS2 data. 

 Figure~4g begins regions in the outer half of the accretion disk. The `stream' here
 agrees well with the `back', but departs increasingly in the successive regions. In Figure~4g there still
 is good agreement between the `stream' and 'back'.  The Table~4 annulus \# 23 radial position fits that of
 the observational data. The synthetic spectrum, with $T_{\rm eff}=9855$K, fits the `back' and `stream'
 well shortward of 4000\AA~but has too large intensity on the longward side. As in Figure~4f, a reduction
 of the model annulus $T_{\rm eff}$ of a few hundred Kelvins produces a close fit to the DS1 data
 longward of 4000\AA~but interpolates half way between the `front' and `back' on the shortward side.
 
 Figure~4h presents DS1 data for radial designation 0.40. The `stream'
 intensity now exceeds that from the `back' longward of 4000\AA~but agrees shortward of there. 
 The synthetic spectrum is in good agreement with the prism data shortward of 4000\AA~and continues
 agreement with the `stream' data to 5000\AA. The `front', `back', and `stream' now begin to show
 differences over much of the spectral region covered. 

 The dotted curve in Figure~4h is DS2 annulus `D'. 
 It would compare with prism `.35' data, half way between Figure~4g
 and Figure~4h, if those data were available.
 The discrepancy, in part, is due to observational data originating
 over a finite annular width; the hotter inner edge provides an enhanced contribution.
 Annulus `D' closely agrees with the DS1 data of Figure~4g shortward of 5000\AA~but disagrees with the
 slope of the Figure~4g synthetic spectrum, more closely fitting the Figure~4h synthetic spectrum
 SED while displaced to a larger intensity. Annulus `D' could be plotted either in Figure~4g or
 Figure~4h; for mimimum congestion we have chosen Figure~4h.

 Figure~4i plots DS1 data for radial region 0.50. 
 The corresponding
 Table~4 annulus is \# 37 with standard model $T_{\rm eff}=6985$K (inner edge), represented by the 
 lower synthetic spectrum.
 DS2 annulus `E' (the lower dotted line), 
 fits the lower 
 synthetic spectrum well. 
 The annulus \#~37 segment $0\arcdeg$ to $90\arcdeg$ is
 set to $T_{\rm eff}=7500$K and is represented by the upper synthetic spectrum. 
 The upper dotted line is DS2 annulus `F', corresponding to the same radial region as
 annulus `E' but azimuth $0\arcdeg$ to $90\arcdeg$ (the 'stream'). Since annulus `F' covers a much larger radial
 region than the upper synthetic spectrum we do not regard the difference as serious.
 The (lower) standard model synthetic spectrum agrees well with the `back'; its representative
 point, at log~${\rm r}/{\rm r_{L1}}=-0.3$, Figure~1, falls on the dotted line below the observed data points.
 The higher temperature annular segment falls on the Figure~1 heavy continuous curve passing 
 through the region of
 observed points.

 Figure~4j shows DS1 data for radial region 0.60. The corresponding Table~4 annulus
 is \# 44, $T_{\rm eff}=6158$K (inner edge). The lower heavy line is that model. To represent the `stream' we have 
 reset the annulus \# 44 $T_{\rm eff}$, between azimuth $0\arcdeg$ and $90\arcdeg$, to 8600K. The
 upper synthetic spectrum represents that annular segment. 
 The DS1 `stream' and `back' agree shortward
 of 4000\AA~but the very large `stream' Balmer jump is larger than the synthetic spectrum represents.
 The differences between the `front' and `back' are close to their errors and we do not regard
 their differences or their departure from the synthetic spectrum as significant.
 Note the change in ordinate scales from Figure~4a to Figure~4j; the intensity values change by roughly
 a factor 50.

 We have interpolated $T_{\rm eff}$ values for the azimuth $0\arcdeg$ to $90\arcdeg$ segments at
 annuli between \# 30 and \# 44 where the `stream' differs significantly from the `back'. This completes
 our accretion disk model and prepares a comparison with other data sets.

\subsection{Light curve simulations}

 We simulate $UBV$ light curves via synthetic photometry. We calculated synthetic system spectra at 81
 orbital phases, distributed to cover the variable parts of the light curve with adequate phase resolution.
 The synthetic spectra were based on the intensity version of SYNSPEC, which provides automatic
 wavelength-dependent correction for limb darkening, but converted to flux-based spectra for the final output. 
Each synthetic spectrum was weighted by the pass bands for the V,B, and U standard filters and the products
integrated. The integrated flux in V,B, and U, normalized to the flux maxima, as function of phase constitute the
theoretical light curves. 

 The software represents the bright spot as a rim section of elevated $T_{\rm eff}$ \citep{wood1986,wood1989};
 this is a modification of the (otherwise) isothermal rim (\S3.1).
 The bright spot covers the full rim height; the structure of our present program does not allow
 a variable rim thickness with azimuth.
 The rim semi-thickness, $H$, Table~2, follows from the standard model \citep[eq.5.39]{fkr92}.
 It was necessary to assign the region of elevated temperature to an extended azimuth
 region on the rim, in agreement with \citet{smak1994a}, to fit the \citet{wh1954} observations 
 (see Table~6 for the spot parameters).
 Hydrodynamical 2D models of the stream
 impact \citet{rozy1985,rozy1988} identify two shock waves: (1) a shock on a plane perpendicular to the
 orbital plane, roughly bisecting the
 angle between the stream and the rim and terminating at the upstream edge of the stream, and (2) a 
 shock slightly more inclined to the stream and extending far into the disk. Although the simulation is
 2D, \citet{rozy1985} states that a bow shock will develop, prospectively leading to vertical expansion
 upstream. \citet{livio1986} and \citet{ar1998} perform a 3D simulation and find that material from the 
 stream flows over
 the disk if cooling is efficient, applicable to low $\dot{M}$ cases, and is more like an explosion in
 high $\dot{M}$ cases, leading to a bulge extending along the disk rim. 
 The inclined shock wave plane suggests that a more complex model 
 would provide a better physical representation of the bright
 spot than adopted in this study. HO1994 developed a model of this type in their study of
 OY Car, in contrast to the `painted on' rim model of \citet{wood1989} for the same star. 

 Figure~5 shows the fit to the $V$ light curve, Figure~6 shows the fit to the $B$ light curve, and
 Figure~7 shows the fit to the $U$ light curve with a depth residual. A $U$ eclipse depth residual,
 differing from the
 good $V$ and $B$ fits, is suggestively familiar from the binary star literature; 
 it results from the poor representation of
 stellar SEDs by a black body over the Balmer discontinuity. That explanation is not possible here
 since the synthetic spectra simulate the Balmer discontinuity. The far edge of the accretion disk remains
 uneclipsed and it is this region that shows extra unmodeled light shortward of the Balmer discontinuity (Figure~4).
 We suggest that it is this unmodeled, uneclipsed light that reduces the observed depth of U eclipse
 below the model.
 These light curves adopt a 20,000K WD as
 discussed in the previous section.	The eclipse depths are more strongly correlated with orbital inclination
 than WD $T_{\rm eff}$
 and so are not useful in distinguishing between a 20,000K WD and a 40,000K WD. 
 We call attention to the downward trend in the Figure~6 residuals between orbital phases 0.2 and 0.8.
 \S6 discusses this feature in terms of a vertically extended rim.
 It is of interest that the model provides a good light curve fit over the range of dates from April~17, 1953
 ($B$)
 to June~13, 1953 ($U$).

 We have tested the sensitivity of our model to our choice of $R_{\rm cutoff}=0.6R_{\rm L1}$ by setting
 $R_{\rm cutoff}=0.7R_{\rm L1}$ and recalculating the model light curves. The eclipse widths remain the
 same to visual detectability limits while the eclipse depths become systematically smaller by small
 amounts; the $U$ calculated depth now fits the observations precisely while the $V$ and $B$ calculated depths are
 slightly too shallow. 
 
 \section{SED fits to observed spectra}

 We apply the model developed in previous sections,  including the rim 
 bright spot from the immediately
 previous section, hereafter the \S4. model, to represent observed spectra. The SYNSPEC 
 synthetic spectra
 for the annuli of Table~3 are produced in both an intensity format and a flux-based format; the latter format
 requires adoption of a wavelength-independent limb darkening coefficient.
 The intensity simulations of \S4 required the intensity-based synthetic spectra. 
 We used the intensity-based synthetic
 spectra, which automatically correct for wavelength-dependent limb darkening, and converted the
 output to flux units for comparison with the observed spectra which are tabulated in physical flux units.

 \subsection{SED fits to $FUSE$ and $FOS$ spectra}

 The $FOS$ spectra have already been used by BA1998
 to produce the MEM data modeled in Figure~4.
 Our objective in this section is to combine the $FOS$ and $FUSE$ spectra and
 model those observed spectra. The $FOS$ and $FUSE$ spectra have exposure times listed in Table~1. 
 $FUSE$ orbit03 exposure starts at orbital phase 
 $0.4333$ and ends at $0.5117$. $FUSE$ orbit04 exposure starts at orbital
 phase $0.7863$ and ends at orbital phase $0.8750$.	Both of the $FUSE$ exposures
 start and end outside eclipse. 
 The first G160L exposure, data set Y2AH0201T, starts at
 orbital phase $-0.090$ and ends at orbital phase $0.130$. The second G160L exposure, data set
 Y2AH0401T, starts at orbital phase $-0.063$ and ends at orbital phase $0.156$.
 As BA1998 indicate, in each case 
 the $FOS$ data set consists of 691 spectra produced in `rapid readout' mode.
 In each $FOS$ case, our spectrum is the sum of the exposures in the data set. 

 To simulate the sum of the exposures for the $FOS$ spectra, we calculated 33 synthetic system spectra
 equally spaced in phase between phase -0.063 and 0.156, corresponding to the Y2AH0401T data set.
 These spectra all used a 20,000K model for the WD, specifically for 
 $\dot{M}=5.0{\times}10^{-9}M_{\odot}~{\rm yr}^{-1}$. BA1998 note that there was a $70\%$ reduction
 in $\dot{M}$ from the DS1 observations to the $FOS$ observations. Experiment shows that a change
 like this does not make a large change in the calculated spectral gradient; the major change is in the flux
 level. Rather than calculate a complete new set of TLUSTY annuli and a corresponding new BINSYN
 model, we preserve the adopted $\dot{M}$ and subsume the effect of the $\dot{M}$ change in the
 normalization factor to superpose the model on the observed spectra (see below).
 We summed the spectra and divided by 33 to produce an average spectrum through eclipse. Comparison
 of the average spectrum with one that is outside eclipse shows that the average spectrum closely
 fits the outside-eclipse profile but has a flux level that is $88\%$ of the outside-eclipse
 spectrum. We use this average spectrum for comparison with both of the $FOS$ spectra. 
 
 Figure~8 compares two synthetic spectra with the YAH0201T spectrum. The lower synthetic spectrum is the
 average spectrum described above and the upper synthetic spectrum is the 40,000K model. It is apparent
 that the synthetic spectrum departs significantly from the observed spectrum. 
 In the following
 section we will find that the same is true of the $IUE$ spectra and that the SED is appreciably variable,
 temporally, with $T_{\rm eff}$ profiles that differ from the standard model.  
 A fair overall fit can be achieved 
 shortward of 2100\AA~with a normalizing factor of
 $7.5{\times}10^{41}$, corresponding to a distance of 281pc. 
 This distance is too large because of the failure to allow for the reduced $\dot{M}$ drescribed above.
 A correction moves the calculated distance toward the adopted 250pc.
 The 40,000K WD model shown (calculated for an outside-eclipse phase), with the same normalizing factor, 
 lies well above the observed spectrum and
 would require a larger normalizing factor to fit the observed spectrum. Reducing to $88\%$ of the calculated
 flux to allow for effects of eclipse still leaves a large discrepancy.
  
 A completely self-consistent loop for the \S4. model would determine a normalizing factor that
 reproduces	the assumed initial model distance of 250pc. Note that this analysis, while adopting values
 of $M_{\rm wd}$ and $\dot{M}$, leaves the distance as a parameter to be determined. By contrast, in Figure~4, the
 data to be fit involve an assumed distance but there is no adjustable parameter in the comparison
 with the model. 

 Figure~9 presents the FUV spectrum for the Figure~8 configuration. The $FUSE$ spectrum is outside eclipse so
 the synthetic spectrum for the 20,000K model also represents that condition, as does the 40,000K synthetic
 spectrum. 
 The \S4. model 
 (lower synthetic spectrum) fits
 the $FUSE$ spectrum approximately but does not reproduce the very large number of fairly deep
 absorption features. The absorption features in the 1120\AA~to 1150\AA~region of
 the synthetic spectrum are replicated with much larger amplitude in the $FUSE$ spectrum.
 FR2003 discuss the $FUSE$ spectra in detail.
 Reddening has a strong effect; a change from E(B-V)=0.00 to E(B-V)=0.01
 raises the $FUSE$ spectrum by $17\%$ without a detectable change in the slope.
 The reddening-corrected spectrum fits the synthetic spectrum better but, as noted below, the $FUSE$ spectrum
 shows phase-wise variation that makes the value of the fit questionable. We do not consider the improved
 fit necessarily as support for a value of E(B-V)=0.01. 
 Note that the emission excess shortward of 965\AA, described by
 FR2003, is clearly present. Also note the blueward displacement of the Ly$\beta$ and Ly$\gamma$
 absorption features; the model approximately reproduces their depths. The correponding Doppler
 shift is ${\sim}3000~{\rm km}~{\rm s}^{-1}$.

 Figure~10 combines the G160L spectrum from the Y2AH0401T set with the $FUSE$ orbit04 spectrum
 (the phase of the $FUSE$ spectrum is unrelated to the phase of the G160L spectrum;
 they have been plotted together for economy--otherwise separate plots would be required).
 Because of the same total phase range for the two $FOS$ spectra and the same number of contributing
 individual spectra they
 would be expected to be essentially identical.
 Yet the G160L spectrum
 shows an overall flux reduction,
 is substantially fainter in the 1200\AA~to 1500\AA~region, but now shows an excellent fit longward
 of 2000\AA. The accretion disk SED has changed between the times of the two observation sets.

 Figure~11, which shows the FUV part of Figure~10, poses a problem.	As with Figure~9, the 20,000K
 model represents an outside-eclipse phase. The $FUSE$ spectrum is outside eclipse
 so it would be expected
 to show little difference from Figure~9, yet the difference is striking. Not only is the flux level
 lower, there appear to be absorption bands that are not prominent in Figure~9.
 Note the broader, deeper and more complex Ly$\beta$ and Ly$\gamma$ features, but the emission excess
 shortward of 965\AA~is unaffected.	The $FUSE$ spectrum shows variation which may be both/either
 temporal or phase-dependent.

 \subsection{SED fits to $IUE$ spectra}

 The $IUE$ archive lists 31 LWP and LWR	spectra, and 37 SWP spectra, mostly
 observed in 1980 and mostly taken in pairs including a LWR exposure and a SWP exposure.
 We used the BA1995
 ephemeris  $T_{\rm min}= {\rm HJD}~2443904.87872+0.196671278E$ to calculate orbital
 phases. 
 The UX UMa orbital period is short enough, and the $IUE$ exposure time long enough, that the second 
 exposure of a pair occurred at an
 orbital phase differing appreciably from the first. For example, in the pair SWP10128+LWR08799
 the first exposure started at orbital phase $0.2887$ while the second started at 
 $0.4069$. In cases
 where both exposures are outside eclipse 
 the phase displacement between exposures makes an 
 undetectable change in the synthetic spectra; consequently we combine the observed spectra of a
 pair for analysis. Separate tests with outside-eclipse synthetic spectra show nearly undetectable
 phase-wise variation in the SED even though the model includes the rim bright spot.
 The observed spectra do show cycle to cycle changes 
 and there are year to year
 changes that are clearly apparent \citep{holm1982} (see below).
 We find that our model with a 20,000K WD provides an appreciably better fit, in all cases tested,
 than a 40,000K WD. A 40,000K WD produces a too-steep spectral gradient (see below).
 
 Figure~12 presents a fit to SWP10371+LWR09051. The orbital phase at the start of the SWP exposure
 was $0.2867$ and the phase at the start of the LWR exposure was $0.1615$. Both 
 exposures were
 outside eclipse. The synthetic spectrum has been divided by $5.2{\times}10^{41}$ for this
 comparison, showing a reasonably close fit to the $IUE$ spectra except beyond 2600\AA;
 the corresponding distance is 234pc. This fit appears to provide support for our adopted distance of 250pc,
 but, as we see in the following discussion, the support is ambiguous.
 The upturn at 2600\AA~is not due to the secondary star since BINSYN includes an explicit model of the 
 secondary star in the simulation; this effect mimics the Figure~8 anomaly.
 Plots of the $IUE$ spectra and the G160L plus prism spectra show a substantially different flux level in
 the two cases, indicating a likely change in $\dot{M}$ between the times of observation.
 (The $IUE$ 1350\AA~continuum flux in Figure~12 is about 3.0 ordinate units	while the corresponding
 level in Figure~8 is about 2.2 ordinate units.)
 
 Figure~13 presents a fit to SWP10128+LWR08798. The orbital phase at the start of the SWP exposure
 was $0.2887$ and the phase at the start of the LWR exposure was $0.4069$.
 The synthetic spectrum has been divided by $5.2{\times}10^{41}$, as in Figure~12. The discrepancy
 beyond 2600\AA~in Figure~12 now afflicts Figure~13 beyond 1700\AA. A possible postulate to explain
 Figure~13 is a lower accretion disk temperature gradient. The required change is drastic: A 20,000K
 WD and a 12,000K isothermal accretion disk provide a good fit, with a normalizing factor of 
 $3.5{\times}10^{41}$, placing the system at a distance of 192pc. But the excess radiation above the 
 synthetic spectrum with the Figure~12 normalizing
 factor, nearly consistent with the adopted distance of UX UMa, indicates an actual {\it increase} in total 
 disk luminosity, contradicting the implication of the
 reduced normalizing divisor. 
 Thus, the isothermal accretion disk model fits
 the $IUE$ spectrum but does not provide a believable system model.

 Finally, Figure~14 presents a fit to SWP10677+LWR09388. This spectrum permits an excellent fit by a
 standard model, but with a mass transfer rate of $\dot{M}=3.0{\times}10^{-9}M_{\odot}~{\rm yr}^{-1}$;
 the corresponding distance is 165pc. The actual mass transfer rate in UX UMa may, at times, equal
 the value quoted, but if the accretion disk reaches equilibrium, the flux values will differ appreciably from the
 Figure~14 values.

 Our conclusions from simulation of these and other $IUE$ spectra are:
 (1) The Figure~8 anomaly occurs frequently but may start at shorter or longer wavelengths than 2000\AA.
 The source of the anomaly is unknown.
 (2) In some cases, the SED in a variable $\dot{M}$ system can be accurately fit by a nonstandard model
 (e.g., isothermal) that does not lead to a believable system model; the fit cannot be used even to
 constrain the disk luminosity.
 (3) As a caution to disk modelers, a good standard model fit in a system that shows spectrum variability
 cannot be taken at face value,
 even if the $\dot{M}$ indicates that the accretion disk is stable against outburst, unless there is independent
 evidence concerning the system distance.

\section{Discussion}

 An important part of the system analysis for UX UMa depends on knowledge of its distance. 
 BA1995 calculate a value of 345pc with an estimated error
 of 34pc. The 345pc determination depends on fits in a color-magnitude diagram, with one wavelength at 1523\AA.
 The fits use a theoretical color-magnitude relation derived on assumed radiation characteristics
 of the source. If the source can be represented by a black body, the derived distance is 401pc, and if
 standard Kurucz model atmospheres, 312pc. We feel the residuals in the SED fits found in the present study are
 large enough that an assumed radiation characteristic for the source should be treated with caution, and that
 the estimated error in the 345pc determination could be larger.
 Other distance determinations are in \S3. The \citet{knigge2006} method is an important improvement on the 
 \citet{bailey1981} relation;
 as cited in \S3., this method leads to a distance of 215pc if the secondary provides all of the system K-flux and
 376pc if the secondary provides 1/3 of the system flux. Based on the parameters of our \S4. model, and adopting a
 secondary $T_{\rm eff}=3575$K \citep[Table~3]{knigge2006} the secondary provides 0.48 of the system flux at
 2.2$\mu$, leading to a calculated distance of 312pc. 
 Consider the sensitivity of the calculated distance to variation of system parameters.
 The flux ratio secondary/(secondary+disk) is sensitive to 
 the accretion disk flux which in
 turn depends on $\dot{M}$. (The contribution of the adopted 20,000K WD is only a few tenths of a percent at 2.2$\mu$.)  
 From Figure~1 we estimate that the scatter of the observational data constrains $\dot{M}$ within a factor of 
 about 2 for the adopted system distance
 of 250pc. 
 Increasing $\dot{M}$ to
 $1.0{\times}10^{-8}M_{\odot}~{\rm yr}^{-1}$ produced a new calculated distance of 332pc. 
 We increased $M_{\rm wd}$ by $10\%$, while maintaining $M_{\rm sec}$ fixed, to test the calculated distance
 sensitivity to $M_{\rm wd}$ variation. The new distance was 323pc.
 Variation in the
 observed K magnitude of a few times 0.01 produces variation of only a few parsecs in the calculated distance.
 We propose a distance of $312\pm30$pc as the best currently 
 available distance determination to UX UMa and note the accordance
 with the value found by BA1995 using a fit to Kurucz model atmospheres.

This study finds that a standard model 
${\dot{M}}=5.0{\times}10^{-9}M_{\odot}~{\rm yr}^{-1}$ accretion disk surrounding a 20,000K (but see below),
$0.47M_{\odot}$ WD at a distance of 250pc provides a model that reasonably fits spectral intensity 
data (Figure~4) and spectral flux
data (Figure~8 through Figure~11). 
All of the DS1 and DS2 data can be equally well represented if the distance is 345pc and
${\dot{M}}=1.0{\times}10^{-8}M_{\odot}~{\rm yr}^{-1}$. 
In particular, the observational data of Figure~1 depend only on the adopted distance. 
Interpolating between the 250pc and 345pc calibrations, for our preferred distance of 312pc,
the corresponding $\dot{M}=8{\times}10^{-9}M_{\odot}~{\rm yr}^{-1}$. We take this value to be our final
result for the mass transfer rate. In principle it would be possible to iterate our solution, starting over
with the new distance and $\dot{M}$ determination but, based on our study of the
${\dot{M}}=1.0{\times}10^{-8}M_{\odot}~{\rm yr}^{-1}$ case at 345pc, we believe there would be no
improvement in any of the plots presented in this paper. 

\citet{puebla2007} (hereafter P2007) use a separate method to study accretion rates and, for UX~UMa 
(their Table~2) find
${\dot{M}}=1.7{\times}10^{-8}M_{\odot}~{\rm yr}^{-1}$ from black body fits and
${\dot{M}}=1.4{\times}10^{-8}M_{\odot}~{\rm yr}^{-1}$ for their model accretion disk fits, adopting a
distance to UX UMa of 340pc and $M_{\rm wd}/M_{\odot}$ in the range 0.4-0.8. This result is in approximate
agreement with our results. 
In their Figure~6 P2007 fit $IUE$ spectra of
UX~UMa with two models. Their second model, for $M_{\rm wd}/M_{\odot}=0.4$, finds
${\dot{M}}=5.5{\times}10^{-9}M_{\odot}~{\rm yr}^{-1}$, in close agreement with our model.
P2007 parameterize the WD contribution with ${\zeta}=f_{\rm wd}/f_{\rm disk}$ and define
a ``disk-dominated'' system as one in which ${\zeta}<0.1$, where the flux values are integrated contributions
from 1500\AA~to 3250\AA. 
We would prefer
to define ``disk-dominated'' in terms of an integration extending to 950\AA~since a large part of the
flux from a hot WD can occur shortward of 1500\AA.
Because of the ambiguity of the WD $T_{\rm eff}$ in UX~UMa
(see below), we
do not estimate a value for $\zeta$ except to note that, from Figure~4a, the accretion disk supplies well
over $90\%$ of the system flux if the WD $T_{\rm eff}=20,000$K (compare with the zero temperature WD where
the accretion disk supplies $100\%$ of the system flux, excluding the secondary). 

Several studies (\S3) support the adopted $M_{\rm sec}=0.47M_{\odot}$. The adoption of $q=1.0$ places the
system at the boundary of instability against dynamical scale mass transfer and a $M_{\rm wd}$ larger than 
$0.47M_{\odot}$ seems likely. \citet{bap1995} state that adoption of the \citet{smak1994a} 
$M_{\rm wd}=0.70M_{\odot}$ increases the system distance from 345pc to 386pc; the effect of a smaller $q$
is to increase the size of the accretion disk and so make it more luminous. 
A WD mass greater than $0.47M_{\odot}$ would be smaller and would produce a deeper potential well so, for the same
$\dot{M}$, the accretion disk would be hotter and the spectral gradient would be steeper. Preservation of a fit to the
calibrated temperatures of Figure~1 would require either a reduction in $\dot{M}$ or a nonstandard model
temperature profile. 

The evidence concerning the WD $T_{\rm eff}$ is ambiguous. BINSYN requires specification of a WD $T_{\rm eff}$
and our adopted \S4. model includes a 20,000K WD
but, as Figure~4a shows, this model differs almost negligibly from a zero temperature WD; the hotter
inner annuli, Table~4, contributions dominate the system synthetic spectrum. 
A 40,000K or hotter WD, directly visible to
the observer, would be strongly
inconsistent with the Figure~3 intensity data while the SED data, Figure~8 through Figure~11, all are 
inconsistent with a
hot WD. The only data supporting a hot WD are Figure~4a and Figure~4b. 
Although the DS1 data plotted in Figure~4a and Figure~4b superficially support a 40,000K WD, we
find it more attractive to attribute the excess flux source to something like a boundary layer in common
with the other unmodeled hot sources described in connection with Figure~4a.
We argue that, in Figure~4a, 
the central region is seen through an absorbing layer, likely the ``transition region''
between the accretion disk and the fast wind \citep{knigge1997}, which produces the deep absorption features.

Studies of other cataclysmic variable systems generally support a WD $T_{\rm eff}$ hotter than 34,000K 
\citep{sion1999,knigge2000,sion2008} at the orbital period of UX UMa.
It is informative that the WD temperature determination for DW UMa obtained during a low state of
accretion when the WD was clearly visible \citep{knigge2000} were much higher than during the normal high
accretion state, indicating that the accretion disk can occult the WD. Since the orbital period of UX UMa 
is outside the range for systems that undergo low states, there is no opportunity to determine how much disk
occultation occurs.
\citet[Table~3]{sion1999} lists CV systems
for which absorbing curtains obscuring the WDs have been calculated. Note that our application would be more
complex since, in addition to obscuring the central region, we require obscuration of the `front' of 
particular annular regions, separate from the WD. 

As we have demonstrated, an accretion disk with constant $\dot{M}$ fits both the DS2 and DS1
observations, and this datum (a fixed, unchanging accretion disk) might
imply that a changed WD explains the difference between Figure~3 and Figure~4b. 
But if $\dot{M}$ has remained 
nearly constant we see no 
mechanism to heat the WD (and we do not believe a changed $\dot{M}$ could heat the WD from 20,000K to
40,000K), while there are documented changes in the accretion disk	(Figure~8 to Figure~13)
and one such change might be formation of a hot emitting region between the times of the two data sets
(June 4, 6, and 7, 1992 for DS2 and Nov. 11, 1994 for DS1).

Our relatively simple bright spot model, \S4.1, represents the $UBV$ light curves of Figure~5 through
Figure~7, including the luminosity maximum just before eclipse. 
These simulations are the basis for our determination of the system orbital inclination. 
UX UMa occasionally shows
anomalous light curves that can not be represented by even a sophisticated model like that
of HO1994. An example is in Figure~15. Less extreme cases show a depressed light level from phase
0.2 to 0.8, with a downward tilt to phase 0.8, followed by a rapid rise to light maximum at phase 0.9. These
features become more pronounced from $V$ to $B$ to $U$. An example is (our) Figure~6 or Figure~3 of \citet{jph1954}.
 \citet{mason1997} attribute this effect to a disk `bulge' upstream from the bright spot and tie it to a 
 similar effect in UV spectra. In Figure~15 we suggest that, except for a temporally enlarged `bulge' 
 or associated disturbance, the system would 
 show a brightness
 peak at orbital phase 0.9 with a brightness reduction to the light level seen at phase 0.15.
 The same `dips' are seen in LMXRBs, discussed by \citet{livio1993} and with the same proposed explanation.
 This subject is discussed further by \citet{knigge1998a} and FR2003. 
 This explanation for anomalous light curves differs from
 the \citet{smak1994b} proposal for circumdisk absorbing
 material. 
 
An important result of this study is that, starting with an adopted distance of 250pc to UX UMa, 
intensity-based observed spectra, flux-based observed spectra,
and photometric data all can be approximately represented by a standard model with 
${\dot{M}}=5.0{\times}10^{-9}M_{\odot}~{\rm yr}^{-1}$ and with an elevated temperature in the accretion
disk quadrant that includes the stream impact region. Application of the \citet{knigge2006} method then
leads to an improved distance estimate of 312pc followed by a revised estimate of the (average) mass transfer rate
(at the times of the DS1 and DS2 data sets). In spite of this success, SED fits show departures from the
standard model
in agreement with BINSYN studies of
IX Vel \citep{linn2007} and QU Car \citep{linnell2008}
which also exhibit departures from a standard model.	
 We postpone attempting to model the IC and the recombination spectrum that fills in the Balmer 
 continuum for the
 `back' spectra, Figure~4c to Figure~4j, to a subsequent publication.  
  
We thank the referee for a prompt report; responding to it substantially improved this paper.
We
are grateful to Dr. Baptista for supplying data from BA1998 for this investigation. 
PG wishes to thank Mario Livio for his kind hospitality at the Space Telescope Science Institute
where part of thi work was carried out.
Support for this work was provided by NASA through grant number 
HST-AR-10657.01-A  to Villanova University (P. Godon) from the Space
Telescope Science Institute, which is operated by the Association of
Universities for Research in Astronomy, Incorporated, under NASA
contact NAS5-26555. 
PS is supported by HST grant GO-09724.06A.

This research was partly based on observations made with the NASA/ESA Hubble Space Telescope, 
obtained at the
Space Telescope Science Institute, which is operated by the Association of Universities 
for Research in Astronomy, Inc. under NASA contract NAS5-26555, and the NASA-CNES-CSA
{\it Far Ultraviolet Explorer}, which is operated for NASA by the Johns Hopkins University
under NASA contract NAS5-32985.

%%% BIBLIOGRAPHY

\clearpage

%%%%%%%%%%%%%%%%%%%%%%%%%%%%%%%%%%%%%%%%%%%%%%%%%%%%%%%%%%%%%%%%%%%
%%% TABLES
%%%%%%%%%%%%%%%%%%%%%%%%%%%%%%%%%%%%%%%%%%%%%%%%%%%%%%%%%%%%%%%%%%%

%%%%%%%%%%%%%%%%%%%%%%%%%%%%%%%%%%%%%%%%%%%%%%%%%%%%%%%%%%%%%%%%%%%

\begin{deluxetable}{lllll}
\tablewidth{0pt} 
\tablecaption{Spectroscopic observations of UX UMa} 
\tablehead{
 \colhead{Date}  & \colhead{Telescope} & \colhead{Exp time} & \colhead{Orb. Phase} & 
 \colhead{Dataset}    \\ 
\colhead{(dd/mm/yyyy)} & \colhead{Instrument} & \colhead{(sec)}  &           
}
\startdata
 14-09-1980	& IUE         &1680      &   0.2887    & SWP10128  \\
 14-09-1980	& IUE         &1800      &   0.4069    & LWR08799  \\
 16-10-1980 & IUE         &1800      &   0.2867    & SWP10371  \\
 16-10-1980 & IUE         &1800      &   0.1615    & LWR09051  \\
 25-11-1980 & IUE         &2699      &   0.3050    & SWP10677  \\
 25-11-1980 & IUE         &1800      &   0.4936    & LWR09388  \\
 03-08-1994 & FOS(G160L)  &3581      &   0.9109    & Y2AH0201T  \\
 03-08-1994 & FOS(G160L)  &3581      &   0.9368    & Y2AH0401T  \\
 24-11-1994 & FOS(PRISM)  &4509      &   0.8383    & Y2AH0601T  \\
 24-11-1994 & FOS(PRISM)  &4509      &   0.8807    & Y2AH0801T  \\
 24-03-2001 & FUSE        &1332(orbit 3)&  0.4333  & B08201010   \\ 
 24-03-2001 & FUSE        &1508(orbit 4)&  0.7863  & B08201010   \\ 
\enddata
\tablecomments{The listings in the orbital phase column refer to the
start of the exposures}
\end{deluxetable}

%%%%%%%%%%%%%%%%%%%%%%%%%%%%%%%%%%%%%%%%%%%%%%%%%%%%%%%%%%%%%%%%%%%
\clearpage

%%%%%%%%%%%%%%%%%%%%%%%%%%%%%%%%%%%%%%%%%%%%%%%%%%%%%%%%%%%%%%%%%%%
\begin{deluxetable}{llll}
\tablewidth{0pt}
\tablenum{2}
\tablecaption{UX UMa Model System Parameters, $M_{\rm wd}=0.47M_{\odot}$}
\tablehead{
\colhead{parameter} & \colhead{value} & \colhead{parameter} & \colhead{value}}
\startdata
${ M}_{\rm wd}$  &  $0.47{M}_{\odot}$	 &$r_s$(pole) &  $0.496R_{\odot}$  \\
${M}_{\rm s}$  &  $0.47{M}_{\odot}$	         &$r_s$(point)   & $0.697R_{\odot}$\\
P    &  0.196671278 days	  & 						$r_s$(side)  & $0.521R_{\odot}$\\
$D$              &  $1.39345R_{\odot}$ & $r_s$back)   & $0.564R_{\odot}$\\	 
${\Omega}_{\rm wd}$         & 85.4 & 	 log $g_s$(pole) & 4.73\\
${\Omega}_s$                &  3.75 & log $g_s$(point) & -4.65\\
{\it i}              &   $70.2{\pm}0.2{\degr}$ & log $g_s$(side)  & 4.65\\
$T_{\rm eff,wd}$         &  $20,000$K\tablenotemark{a}  &log $g_s$(back)  & 4.50\\
$r_{\rm wd}$      &   $0.0165R_{\odot}$ &		 $r_a$ & $0.488R_{\odot}$ \\
log $g_{\rm wd}$  &   7.7&				  $r_b$ & $0.0165R_{\odot}$ \\
  &               &		  $H$    & $0.0158R_{\odot}$  \\

\enddata
\tablenotetext{a}{see text for discussion}
\tablecomments{${\rm wd}$ refers to the WD; $s$ refers to the secondary star.
$D$ is the component separation of centers,
${\Omega}$ is a Roche potential. 
$r_a$ specifies the outer radius 
of the accretion disk, set at the tidal cut-off radius, 
and $r_b$ is the accretion disk inner radius. 
$H$ is 
the semi-height of the accretion disk rim (standard model).}  
\end{deluxetable}

%%%%%%%%%%%%%%%%%%%%%%%%%%%%%%%%%%%%%%%%%%%%%%%%%%%%%%%%%%%%%%%%%%%
\clearpage

%%%%%%%%%%%%%%%%%%%%%%%%%%%%%%%%%%%%%%%%%%%%%%%%%%%%%%%%%%%%%%%%%%%

\begin{deluxetable}{rrrrrrr}
\tablewidth{0pt}
\tablenum{3}
\tablecaption{Properties of accretion disk with mass transfer rate 
$\dot{M}=5.0{\times}10^{-9}~{M}_{\odot}{\rm yr}^{-1}$ and WD mass of $0.47{M}_{\odot}$.}
\tablehead{	  
\colhead{$r/r_{\rm wd,0}$} & \colhead{$T_{\rm eff}$} & \colhead{$m_0$} 
& \colhead{$T_c$} & \colhead{log~$g$}
& \colhead{$z_H$} & \colhead{{$\tau_{\rm Ross}$}}}
\startdata
1.36  &  39724  &  1.256E4   & 4.341E5  &  6.55   & 1.35E8   & 3.48E4\\
2.00  &	 35617  &  1.534E4	 & 4.081E5  &  6.29   & 2.34E8   & 4.18E4\\
3.00  &	 28801  &  1.499E4	 & 3.504E5  &  6.00   & 4.08E8   & 5.25E4\\
4.00  &	 24207  &  1.395E4	 & 3.102E5  &  5.78   & 5.86E8   & 6.40E4\\
5.00  &	 20997  &  1.296E4	 & 2.815E5  &  5.61   & 7.71E8   & 7.63E4\\
6.00  &	 18628  &  1.210E4	 & 2.602E5  &  5.47   & 9.67E8   & 8.95E4\\
7.00  &	 16803  &  1.137E4	 & 2.431E5  &  5.35   & 1.16E9   & 1.03E5\\
8.00  &	 15349  &  1.074E4	 & 2.291E5  &  5.25   & 1.37E9   & 1.17E5\\
9.00  &	 14159  &  1.019E4	 & 2.167E5  &  5.16   & 1.57E9   & 1.29E5\\
10.00 &	 13167  &  9.714E3	 & 2.058E5  &  5.07   & 1.79E9   & 1.41E5\\
12.00 &	 11598  &  8.916E3	 & 1.857E5  &  4.93   & 2.21E9   & 1.57E5\\
14.00 &	 10409  &  8.274E3	 & 1.622E5  &  4.79   & 2.55E9   & 1.54E5\\
16.00 &	  9472  &  7.745E3	 & 1.466E5  &  4.67   & 2.84E9   & 1.51E5\\
20.00 &	  8082  &  6.916E3	 & 1.233E5  &  4.65   & 6.97E9   & 1.39E5\\
24.00 &	  7093  &  6.929E3	 & 1.076E5  &  4.35   & 4.77E9   & 1.32E5\\
30.00 &	  6040  &  5.593E3	 & 0.936E5  &  4.13   & 5.51E9   & 1.42E5\\
\enddata
\tablecomments{Each line in the table represents a separate annulus.
A \citet{ss1973} viscosity parameter $\alpha=0.1$ was used in calculating all annuli.
$r_{\rm wd,0}$ is the radius of a zero temperature, carbon, Hamada-Salpeter WD.
See the text for additional comments.}		 
\end{deluxetable}

%%%%%%%%%%%%%%%%%%%%%%%%%%%%%%%%%%%%%%%%%%%%%%%%%%%%%%%%%%%%%%%%%%%

\clearpage
%%%%%%%%%%%%%%%%%%%%%%%%%%%%%%%%%%%%%%%%%%%%%%%%%%%%%%%%%%%%%%%%%%%

\begin{deluxetable}{crrrcrrr}
\tablewidth{0pt}
\tablenum{4}
\tablecaption{BINSYN model accretion disk with mass transfer rate 
$\dot{M}=5.0{\times}10^{-9}~{M}_{\odot}{\rm yr}^{-1}$ and WD mass of $0.47{M}_{\odot}$.}
\tablehead{	  
\colhead{ordinal val.} & \colhead{${\rm r/r_{\rm wd}}$}  & \colhead{${\rm r/r_{L1}}$} & 
\colhead{$T_{\rm eff}$} 
& \colhead{ordinal val.} &  \colhead{${\rm r/r_{\rm wd}}$}  & \colhead{${\rm r/r_{L1}}$}
& \colhead{$T_{\rm eff}$} 
}
\startdata
 1  &    1.0000  &   .02370  & 33896.  &  24   & 13.3394 & .31610 & 9557.\\
 2  &	 1.1806  &   .02798	 & 33896.  &  25   & 13.9098 & .32962 & 9280.\\
 3  &	 1.3611  &   .03225	 & 35256.  &  26   & 14.4802 & .34313 & 9021.\\
 4  &	 1.9315  &   .04577	 & 32098.  &  27   & 15.0506 & .35665 & 8778.\\
 5  &	 2.5019  &   .05929	 & 28289.  &  28   & 15.6210 & .37017 & 8550.\\
 6  &	 3.0723  &   .07280	 & 25210.  &  29   & 16.1914 & .38368 & 8336.\\
 7  &	 3.6427  &   .08632	 & 22765.  &  30   & 16.7618 & .39720 & 8134.\\
 8  &	 4.2131  &   .09984	 & 20795.  &  31   & 17.3322 & .41072 & 7943.\\
 9  &	 4.7835  &   .11353	 & 19177.  &  32   & 17.9026 & .42423 & 7762.\\
10  &	 5.3539  &   .12687	 & 17823.  &  33   & 18.4730 & .43775 & 7591.\\
11  &	 5.9243	 &	 .14039	 & 16673.  &  34   & 19.0434 & .45126 & 7428.\\
12  &	 6.4947  &   .15390	 & 15683.  &  35   & 19.6138 & .46478 & 7273.\\
13  &    7.0651	 &	 .16742	 & 14820.  &  36   & 20.1842 & .47830 & 7126.\\
14  &	 7.6355  &   .18094	 & 14061.  &  37   & 20.7546 & .49181 & 6985.\\
15  &	 8.2059	 &	 .19445	 & 13388.  &  38   & 21.3250 & .50533 & 6851.\\
16  &	 8.7763  &   .20797	 & 12786.  &  39   & 21.8954 & .51885 & 6723.\\
17  &	 9.3467	 &	 .22149	 & 12244.  &  40   & 22.4658 & .53236 & 6600.\\
18  &	 9.9171	 &	 .23500	 & 11753.  &  41   & 23.0362 & .54588 & 6483.\\
19  &	10.4878	 &	 .24852	 & 11307.  &  42   & 23.6066 & .55940 & 6370.\\
20  &	11.0578  &   .26203	 & 10898.  &  43   & 24.1769 & .57291 & 6262.\\
21  &	11.6282	 &	 .27555	 & 10522.  &  44   & 24.7473 & .58643 & 6158.\\
22  &	12.1986  &   .28907	 & 10176.  &  45   & 25.3177 & .59994 & 6058.\\
23  &	12.7690  &   .30258	 &  9855.     \\
\enddata
\end{deluxetable}

%%%%%%%%%%%%%%%%%%%%%%%%%%%%%%%%%%%%%%%%%%%%%%%%%%%%%%%%%%%%%%%%%%%

\clearpage
%%%%%%%%%%%%%%%%%%%%%%%%%%%%%%%%%%%%%%%%%%%%%%%%%%%%%%%%%%%%%%%%%%%
\begin{deluxetable}{cccccc}
\tabletypesize{\scriptsize}
\tablewidth{0pt}
\tablenum{5}
\tablecaption{Annulus designations and nomenclature for DS1 and DS2}
\tablehead{
\colhead{Data set} & \colhead{spectrograph} & \colhead{annulus} & \colhead{azimuth}	&
\colhead{radial} & \colhead{radial}\\
    &        &      \colhead{subdivision} &  \colhead{range}  & \colhead{designation}
	&  \colhead{range($r_{\rm L1}$)}
}
\startdata
DS1  &  PRISM  & center  &  $0{\arcdeg}$-$360{\arcdeg}$ & --- & 0.000-0.075\\
 &      PRISM  &      stream  &  $0{\arcdeg}$-$90{\arcdeg}$  & 0.10 & 0.075-0.125\\                   
 &      &       back    &  $90{\arcdeg}$-$270{\arcdeg}$ & 0.10	& 0.075-0.125\\
 &      &       front   &  $270{\arcdeg}$-$360{\arcdeg}$ & 0.10 & 0.075-0.125\\
 &      PRISM  &      stream  &  $0{\arcdeg}$-$90{\arcdeg}$  & 0.15 & 0.125-0.175\\                   
 &      &       back    &  $90{\arcdeg}$-$270{\arcdeg}$ & 0.15	& 0.125-0.175\\
 &      &       front   &  $270{\arcdeg}$-$360{\arcdeg}$ & 0.15 & 0.125-0.175\\
 &      PRISM  &      stream  &  $0{\arcdeg}$-$90{\arcdeg}$  & 0.20 & 0.175-0.225\\                   
 &      &       back    &  $90{\arcdeg}$-$270{\arcdeg}$ & 0.20	& 0.175-0.225\\
 &      &       front   &  $270{\arcdeg}$-$360{\arcdeg}$ & 0.20 & 0.175-0.225\\
 &      PRISM  &      stream  &  $0{\arcdeg}$-$90{\arcdeg}$  & 0.25 & 0.225-0.275\\                   
 &      &       back    &  $90{\arcdeg}$-$270{\arcdeg}$ & 0.25	& 0.225-0.275\\
 &      &       front   &  $270{\arcdeg}$-$360{\arcdeg}$ & 0.25 & 0.225-0.275\\
 &      PRISM  &      stream  &  $0{\arcdeg}$-$90{\arcdeg}$  & 0.30 & 0.275-0.325\\                   
 &      &       back    &  $90{\arcdeg}$-$270{\arcdeg}$ & 0.30	& 0.275-0.325\\
 &      &       front   &  $270{\arcdeg}$-$360{\arcdeg}$ & 0.30 & 0.275-0.325\\
 &      PRISM  &      stream  &  $0{\arcdeg}$-$90{\arcdeg}$  & 0.40 & 0.375-0.425\\                   
 &      &       back    &  $90{\arcdeg}$-$270{\arcdeg}$ & 0.40	& 0.375-0.425\\
 &      &       front   &  $270{\arcdeg}$-$360{\arcdeg}$ & 0.40 & 0.375-0.425\\
 &      PRISM  &      stream  &  $0{\arcdeg}$-$90{\arcdeg}$  & 0.50 & 0.475-0.525\\                   
 &      &       back    &  $90{\arcdeg}$-$270{\arcdeg}$ & 0.50	& 0.475-0.525\\
 &      &       front   &  $270{\arcdeg}$-$360{\arcdeg}$ & 0.50 & 0.475-0.525\\
 &      PRISM  &      stream  &  $0{\arcdeg}$-$90{\arcdeg}$  & 0.60 & 0.575-0.625\\                   
 &      &       back    &  $90{\arcdeg}$-$270{\arcdeg}$ & 0.60	& 0.575-0.625\\
 &      &       front   &  $270{\arcdeg}$-$360{\arcdeg}$ & 0.60 & 0.575-0.625\\
\\
DS2  &     &      &    $0{\arcdeg}$-$360{\arcdeg}$ & A & 0.000-0.100\\
     &     &      &	   $0{\arcdeg}$-$360{\arcdeg}$ & B & 0.100-0.200\\
     &     &      &	   $0{\arcdeg}$-$360{\arcdeg}$ & C & 0.200-0.300\\
DS2     &     &      &	   $0{\arcdeg}$-$360{\arcdeg}$ & D & 0.300-0.400\\
     &     &      &	   $90{\arcdeg}$-$360{\arcdeg}$ & E & 0.400-0.700\\
     &     &      &	   $0{\arcdeg}$-$90{\arcdeg}$ & F & 0.400-0.700\\
\enddata
%\tablecomments{}
\end{deluxetable}

%%%%%%%%%%%%%%%%%%%%%%%%%%%%%%%%%%%%%%%%%%%%%%%%%%%%%%%%%%%%%%%%%%%

\clearpage
%%%%%%%%%%%%%%%%%%%%%%%%%%%%%%%%%%%%%%%%%%%%%%%%%%%%%%%%%%%%%%%%%%%

\begin{deluxetable}{lrlrlr}
\tablewidth{0pt}
\tablenum{6}
\tablecaption{Properties of accretion disk bright spot
}
\tablehead{	  
\colhead{parameter} & \colhead{value} & \colhead{parameter} & \colhead{value}
& \colhead{parameter} & \colhead{value}
} 
\startdata
$T_{\rm eff,rim}$  &  6058K	&	HSAZ	&	20 	&	HWDW	&	55\\
HSTEMP  &	 14,000K	&	HSDWD	&	1	&	TDWND	&	6500K\\
HSUP  &	  1	&	TUPD	&	6500K\\
\enddata
\tablecomments{
HSAZ is the position angle (degrees) of the center of the bright spot region, measured
from the line of centers in the direction of accretion disk rotation.
HWDW is the angular width of the bright spot, in degrees.
HSTEMP is the temperature of the bright spot.
HSDWD is the angular width (degrees) of the downwind region from the end of the bright spot
to the merge point with the rim proper.
TDWND is the temperature of the middle of region HSDWD.
HSUP is the angular width (degrees) of the upwind region comparable to HSDWD.
TUPD is the temperature of the middle of region HSUP.
}		 
\end{deluxetable}

%%%%%%%%%%%%%%%%%%%%%%%%%%%%%%%%%%%%%%%%%%%%%%%%%%%%%%%%%%%%%%%%%%%

\clearpage

%%%%%%%%%%%%%%%%%%%%%%%%%%%%%%%%%%%%%%%%%%%%%%%%%%%%%%%%%%%%%%%%%%%
%%% FIGURES
%%%%%%%%%%%%%%%%%%%%%%%%%%%%%%%%%%%%%%%%%%%%%%%%%%%%%%%%%%%%%%%%%%%
\begin{figure}[tb]
\figurenum{1}
\epsscale{0.97}
\plotone{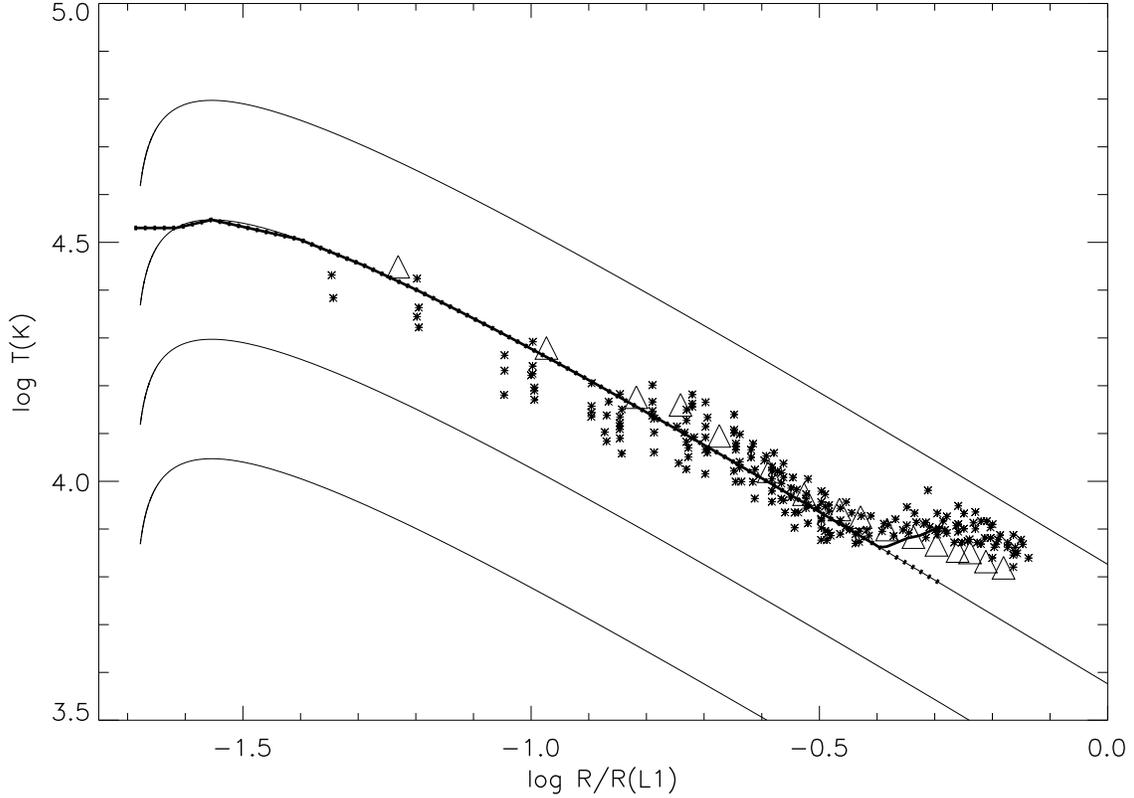}
\epsscale{1.00}
\vspace{0.5cm}
\figcaption{The continuous line is the $T_{\rm eff}$ profile
for the system model, with restricted annular azimuthal sections in
the outer part of the accretion disk. The dotted line is the same
profile but for the remainder of the azimuthal annular sections in the
outer part of the accretion disk.
Crosses mark plotted points from 
\citet[Figure~4a]{ru1992}. Triangles mark `back' side temperatures from
\citet{bap1998} for November 1994 transformed to our adopted distance of 250pc.
Continuous curves, from top to bottom,
are standard models for $\dot{M}/M_{\odot}({\rm yr}^{-1})=5.0{\times}10^{-8}, 5.0{\times}10^{-9},
5.0{\times}10^{-10}, {\rm and}~5.0{\times}10^{-11}$. See the text for details.
\label{F1}}
\end{figure}
%%%%%%%%%%%%%%%%%%%%%%%%%%%%%%%%%%%%%%%%%%%%%%%%%%%%%%%%%%%%%%%%%%%

%%%%%%%%%%%%%%%%%%%%%%%%%%%%%%%%%%%%%%%%%%%%%%%%%%%%%%%%%%%%%%%%%%%
\begin{figure}[tb]
\figurenum{}
\includegraphics[scale=0.5,angle=90]{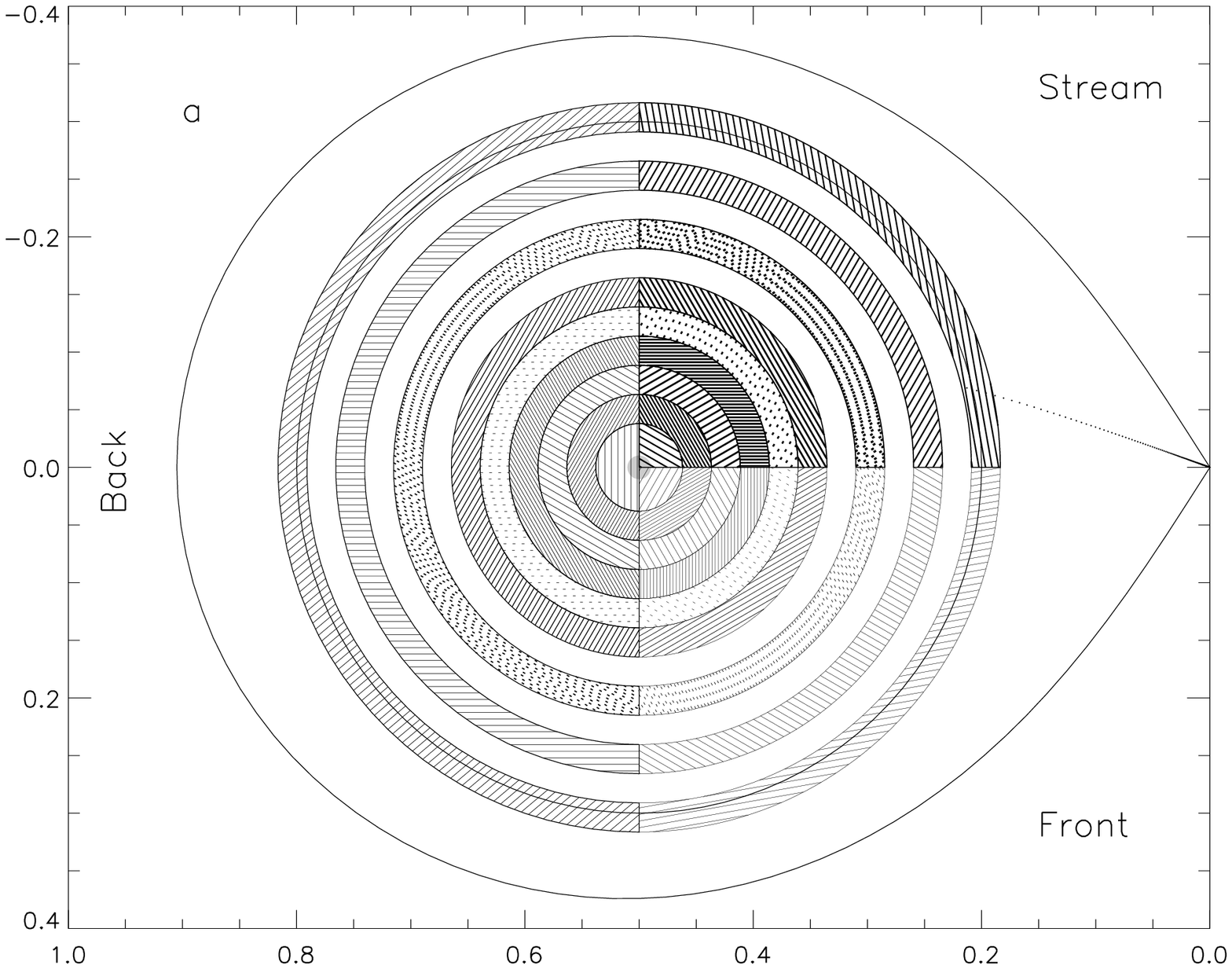}
\vspace{0.5cm}
%\label{F2a}}
\end{figure}
%%%%%%%%%%%%%%%%%%%%%%%%%%%%%%%%%%%%%%%%%%%%%%%%%%%%%%%%%%%%%%%%%%%

%%%%%%%%%%%%%%%%%%%%%%%%%%%%%%%%%%%%%%%%%%%%%%%%%%%%%%%%%%%%%%%%%%%
\begin{figure}[tb]
\figurenum{2}
\includegraphics[scale=0.5,angle=90]{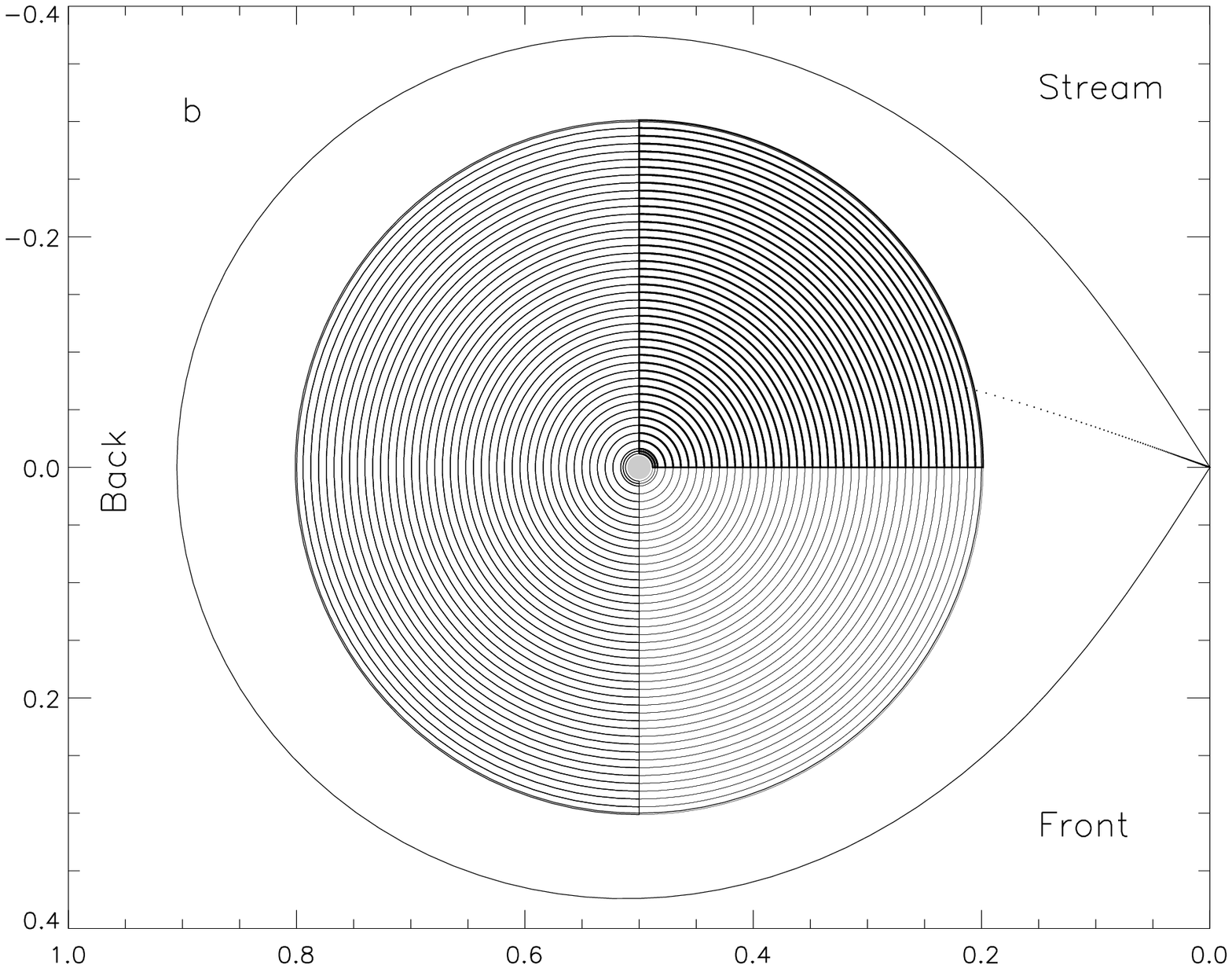}
\figcaption{(a): Segregation of DS1 pixels into concentric rings.
(b): Corresponding BINSYN model with 44 annuli.} 
%\label{F2b}}
\end{figure}
%%%%%%%%%%%%%%%%%%%%%%%%%%%%%%%%%%%%%%%%%%%%%%%%%%%%%%%%%%%%%%%%%%%

%%%%%%%%%%%%%%%%%%%%%%%%%%%%%%%%%%%%%%%%%%%%%%%%%%%%%%%%%%%%%%%%%%
\begin{figure}[tb]
\figurenum{3}
\epsscale{0.97}
\plotone{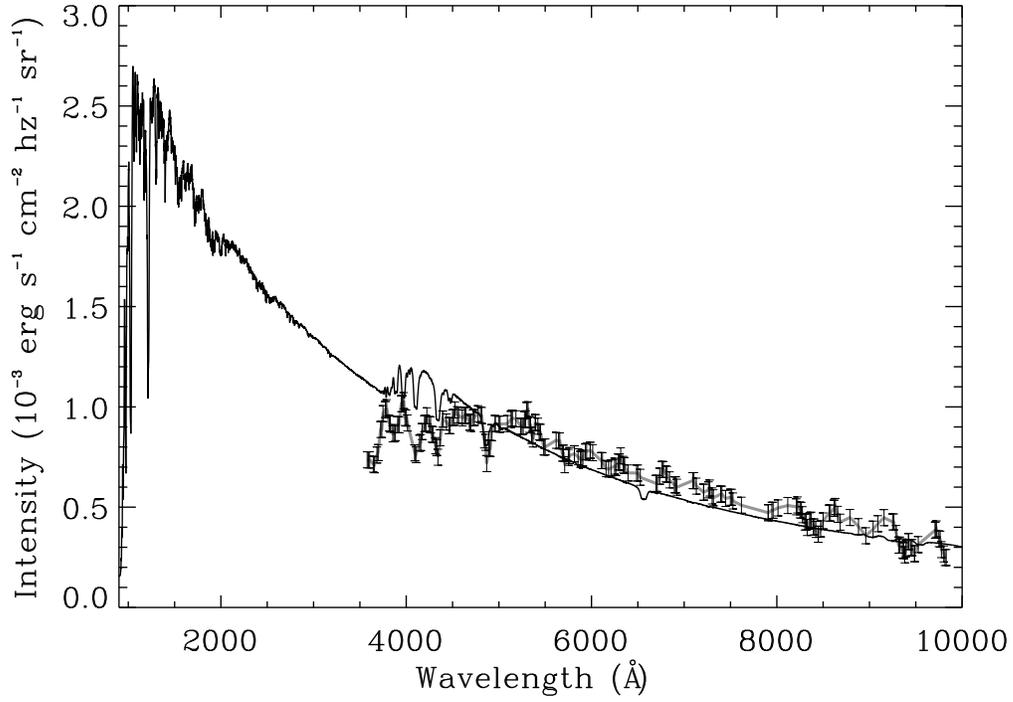}
\epsscale{1.00}
\vspace{0.5cm}
\figcaption{Fit to \citet{ru1994} central region (grey line, region A, Table~5).
The continuous line is a 20,000K WD plus $\dot{M}=5.0{\times}10^{-9}M_{\odot}{\rm yr}^{-1}$
standard model accretion disk. The accretion disk region corresponding to the
observational data extends to 
$0.10{\rm r_{L1}}$, as for the model data.	Except for the adopted 
distance used to produce the data points there are no free parameters in this fit.
See the text for details. 
\label{F3}}
\end{figure}
%%%%%%%%%%%%%%%%%%%%%%%%%%%%%%%%%%%%%%%%%%%%%%%%%%%%%%%%%%%%%%%%%%%

%%%%%%%%%%%%%%%%%%%%%%%%%%%%%%%%%%%%%%%%%%%%%%%%%%%%%%%%%%%%%%%%%%%
\begin{figure}[tb]
\figurenum{4a-4d}
\epsscale{0.97}
\plotone{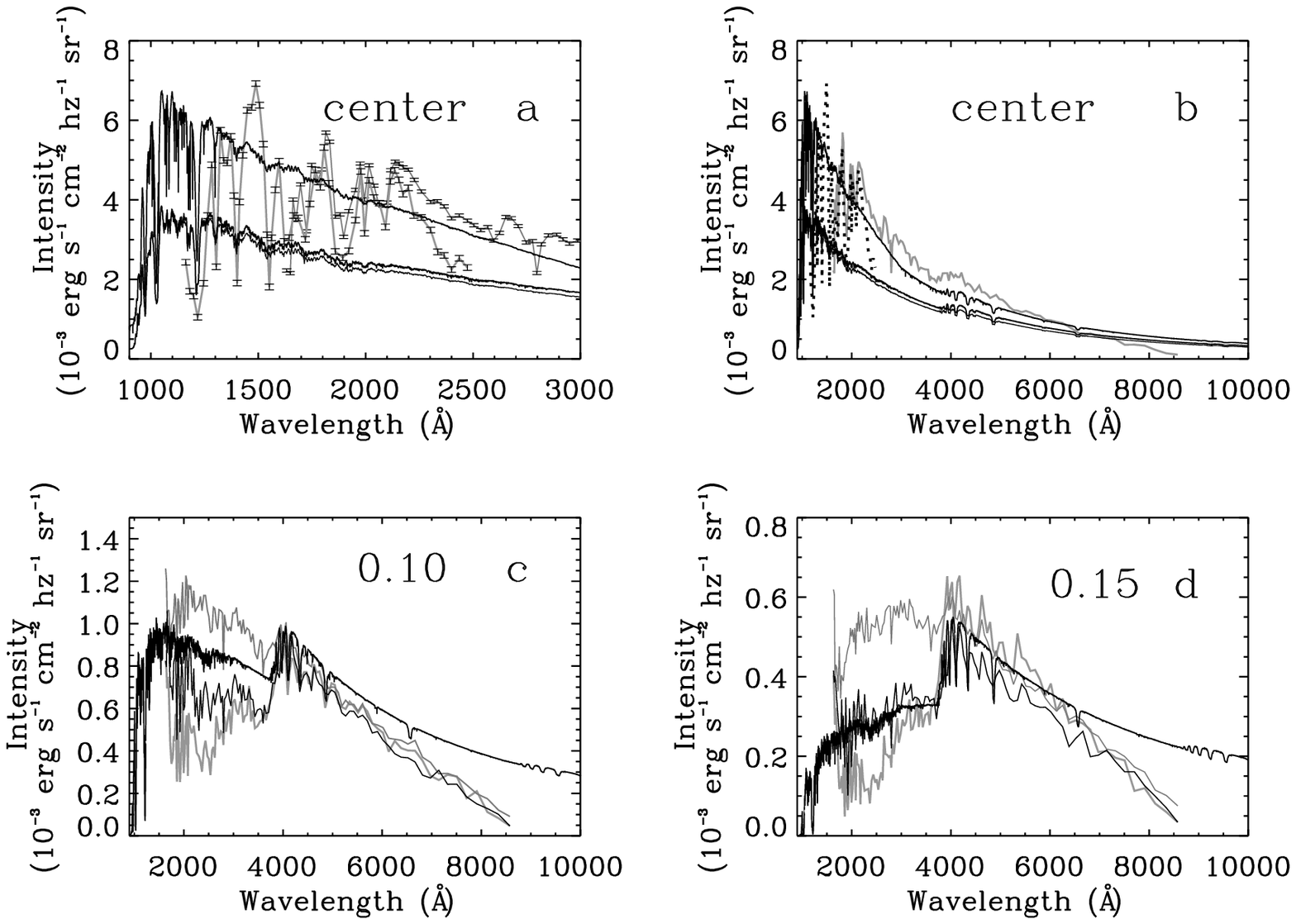}
\epsscale{1.00}
\vspace{0.5cm}
%\figcaption{
%\label{F4a_d}}
\end{figure}
%%%%%%%%%%%%%%%%%%%%%%%%%%%%%%%%%%%%%%%%%%%%%%%%%%%%%%%%%%%%%%%%%%%

%%%%%%%%%%%%%%%%%%%%%%%%%%%%%%%%%%%%%%%%%%%%%%%%%%%%%%%%%%%%%%%%%%%
\begin{figure}[tb]
\figurenum{4e-4h}
\epsscale{0.97}
\plotone{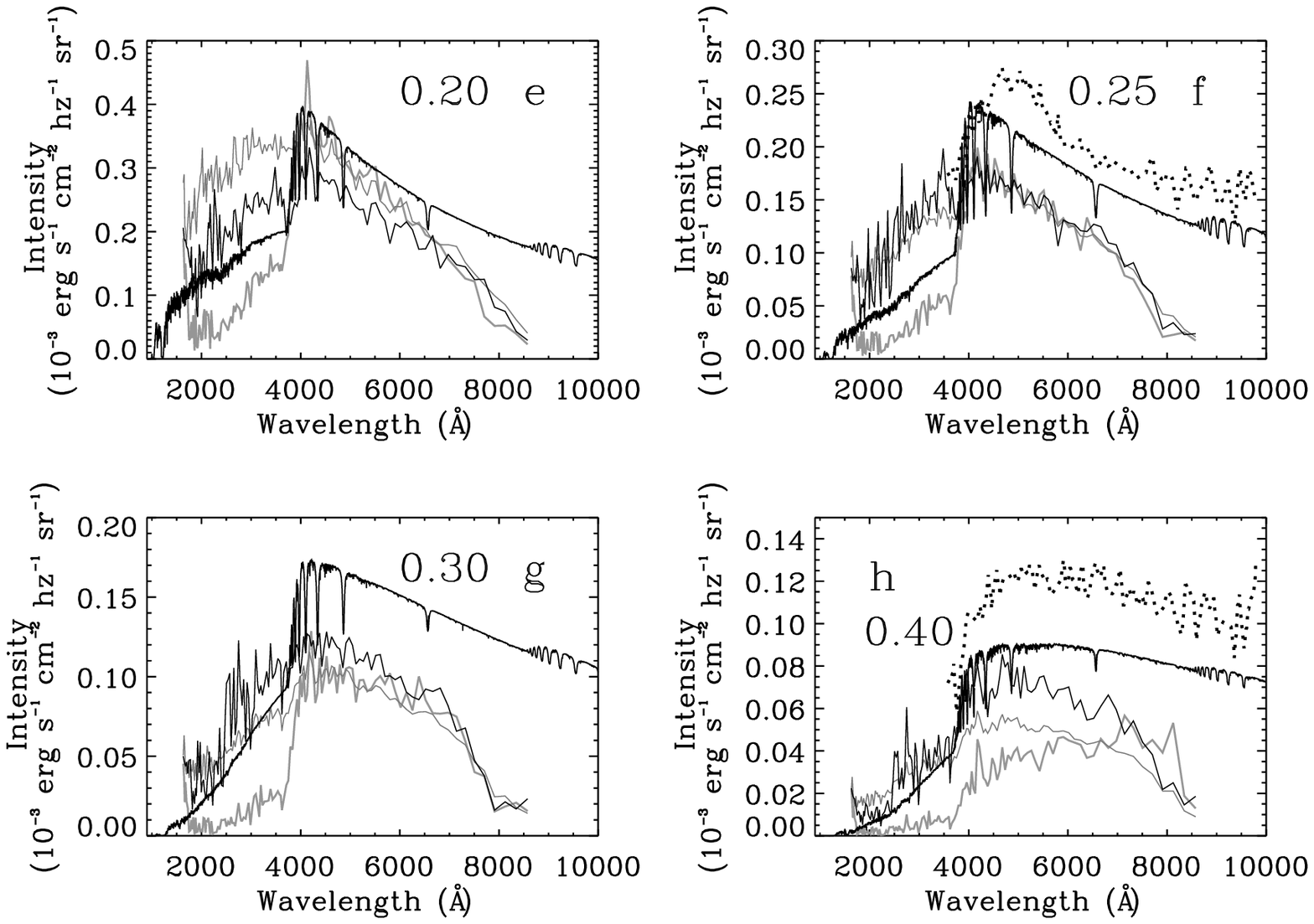}
\epsscale{1.00}
\vspace{0.5cm}
%\figcaption{
%\label{F4e_h}}
\end{figure}
%%%%%%%%%%%%%%%%%%%%%%%%%%%%%%%%%%%%%%%%%%%%%%%%%%%%%%%%%%%%%%%%%%%

%%%%%%%%%%%%%%%%%%%%%%%%%%%%%%%%%%%%%%%%%%%%%%%%%%%%%%%%%%%%%%%%%%%
\begin{figure}[tb]
\figurenum{4a-4j}
\epsscale{0.97}
\plotone{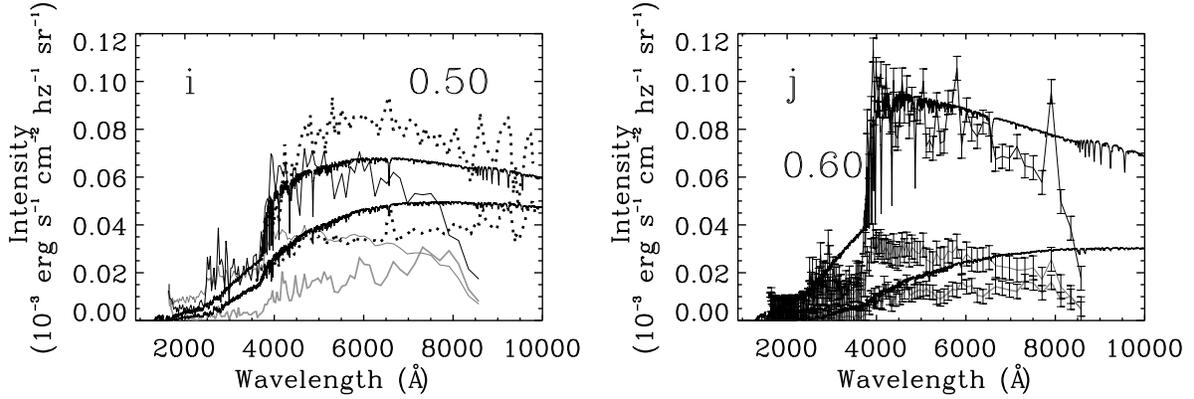}
\epsscale{1.00}
\vspace{0.5cm}
\figcaption{
Fits of synthetic spectra to DS1 and DS2 data. In Figure~4a
and Figure~4b
the broad grey plot is the PRISM and the light grey plot is G160L;
they are the 'center' in Table~5.
In Figure~4c and thereafter the heavy grey line is the 'front',
the light grey line is the 'back', and the light continous line is
the 'stream'. The heavy continuous line is the synthetic spectrum.
In all cases the synthetic spectrum 'front' is the same as the 'back'.
Figure~4c is radial designation (Table~5) 0.10; Figure~4d is radial
designation 0.15. The dotted plot is DS2 radial designation B.
Figure~4e is DS1 radial designation (Table~5) 0.20; Figure~4f
is radial designation 0.25; the dotted line is DS2 radial designation C.
Figure~4g is DS1 radial designation 0.30;
Figure~4h is DS1 radial designation 0.40; the dotted line is DS2 radial
designation D. Note that the 'stream' now is higher than either the 'front' or 'back'.  
Figure~4i is DS1 radial designation (Table~5) 0.50. The lower synthetic
spectrum simulates the 'back' and the upper synthetic spectrum simulates
the 'stream'. The lower dotted line is DS2 radial designation E and the 
upper dotted line is DS2 radial designation F. 
Figure~4j
is DS1 radial designation 0.60. The lower synthetic spectrum simulates
the 'back' and the upper synthetic spectrum simulates the 'stream'.
Note that the DS1 'stream' greatly exceeds either the 'front' or 'back'.
See the text
for details.
\label{F4i_j}}
\end{figure}
%%%%%%%%%%%%%%%%%%%%%%%%%%%%%%%%%%%%%%%%%%%%%%%%%%%%%%%%%%%%%%%%%%%

\clearpage

%%%%%%%%%%%%%%%%%%%%%%%%%%%%%%%%%%%%%%%%%%%%%%%%%%%%%%%%%%%%%%%%%%%
\begin{figure}[tb]
\figurenum{5}
\epsscale{0.97}
\plotone{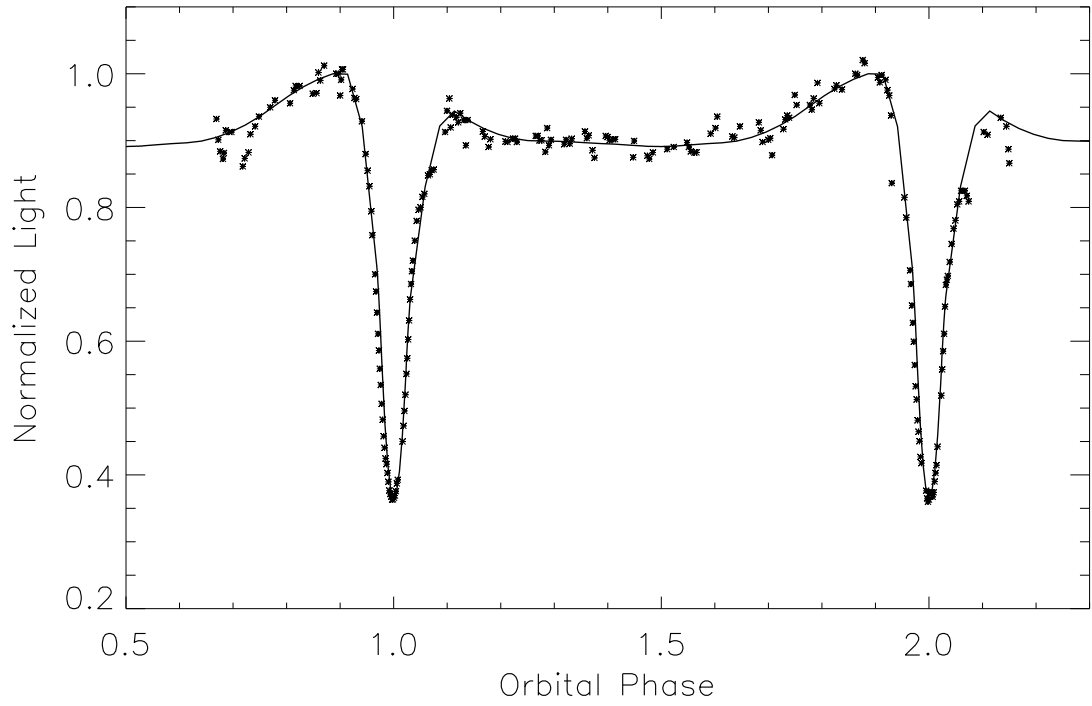}
\epsscale{1.00}
\vspace{0.5cm}
\figcaption{Synthetic light curve fit to V data \citep{wh1954} for
May 21, 1953.
See the text for details.
\label{F5}}
\end{figure}
%%%%%%%%%%%%%%%%%%%%%%%%%%%%%%%%%%%%%%%%%%%%%%%%%%%%%%%%%%%%%%%%%%%

%%%%%%%%%%%%%%%%%%%%%%%%%%%%%%%%%%%%%%%%%%%%%%%%%%%%%%%%%%%%%%%%%%%
\begin{figure}[tb]
\figurenum{6}
\epsscale{0.97}
\plotone{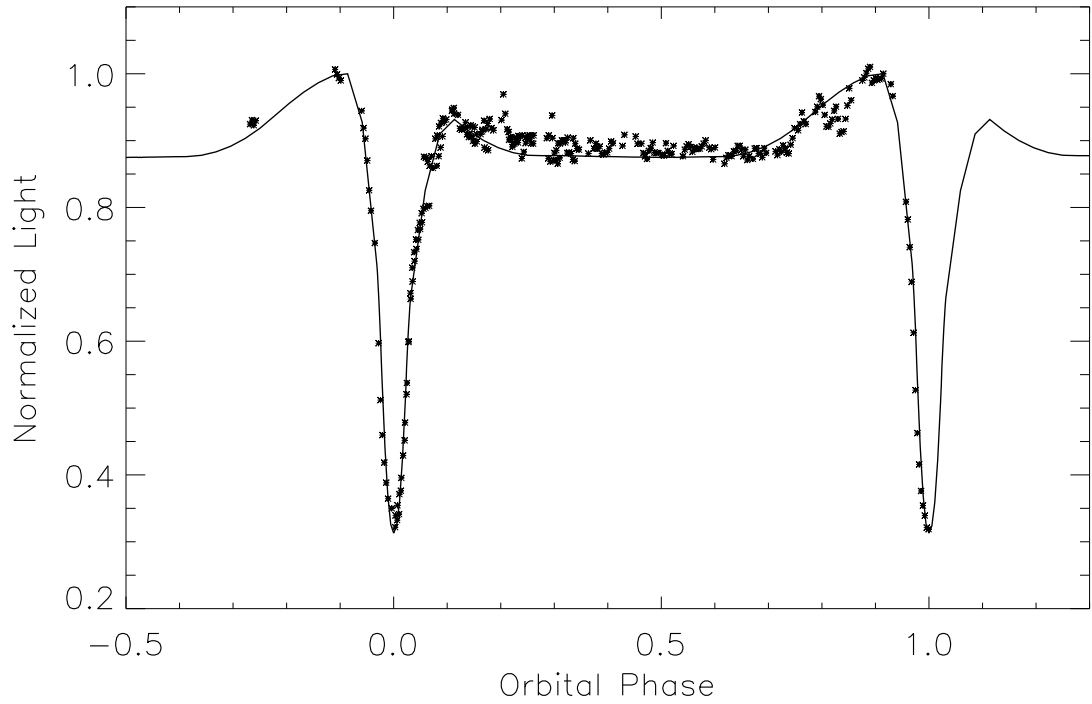}
\epsscale{1.00}
\vspace{0.5cm}
\figcaption{Synthetic light curve fit to B data \citep{wh1954} for
April 17, 1953.	Note the systematic trend in the residuals between orbital
phases 0.2 and 0.8.
See the text for details.
\label{F6}}
\end{figure}
%%%%%%%%%%%%%%%%%%%%%%%%%%%%%%%%%%%%%%%%%%%%%%%%%%%%%%%%%%%%%%%%%%%

%%%%%%%%%%%%%%%%%%%%%%%%%%%%%%%%%%%%%%%%%%%%%%%%%%%%%%%%%%%%%%%%%%%
\begin{figure}[tb]
\epsscale{0.97}
\figurenum{7}
\plotone{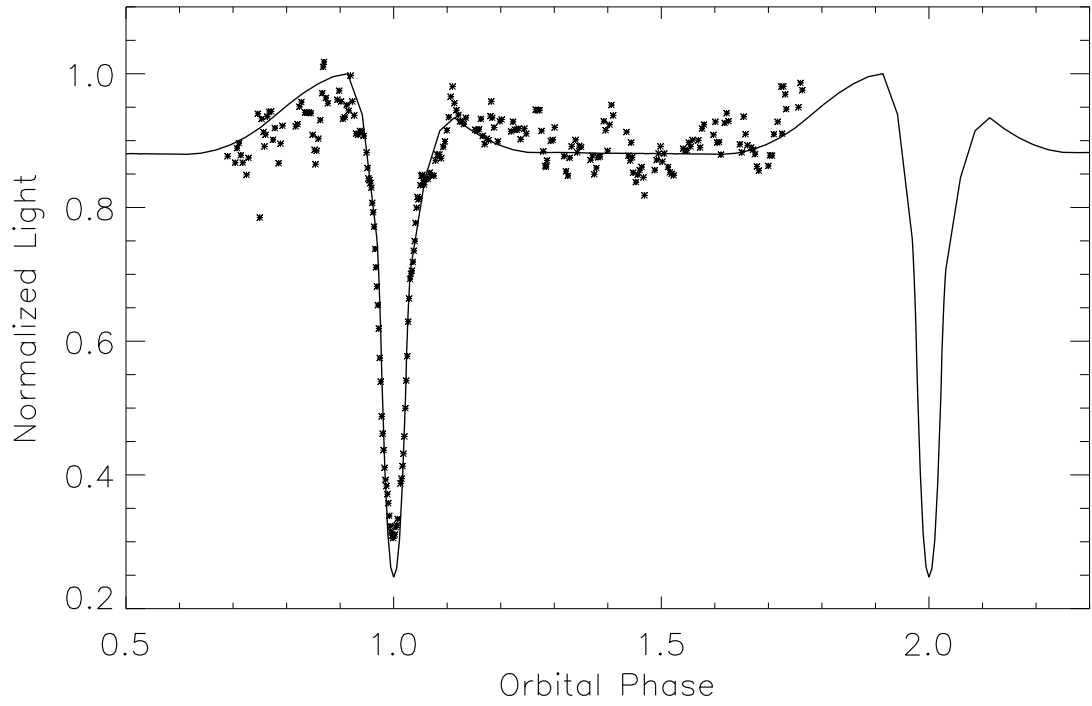}
\epsscale{1.00}
\vspace{0.5cm}
\figcaption{Synthetic light curve fit to U data \citep{wh1954} for
June 13, 1953.
See the text for details.
\label{F7}}
\end{figure}
%%%%%%%%%%%%%%%%%%%%%%%%%%%%%%%%%%%%%%%%%%%%%%%%%%%%%%%%%%%%%%%%%%%
\clearpage

%%%%%%%%%%%%%%%%%%%%%%%%%%%%%%%%%%%%%%%%%%%%%%%%%%%%%%%%%%%%%%%%%%%
\begin{figure}[tb]
\figurenum{8}
\epsscale{0.97}
\plotone{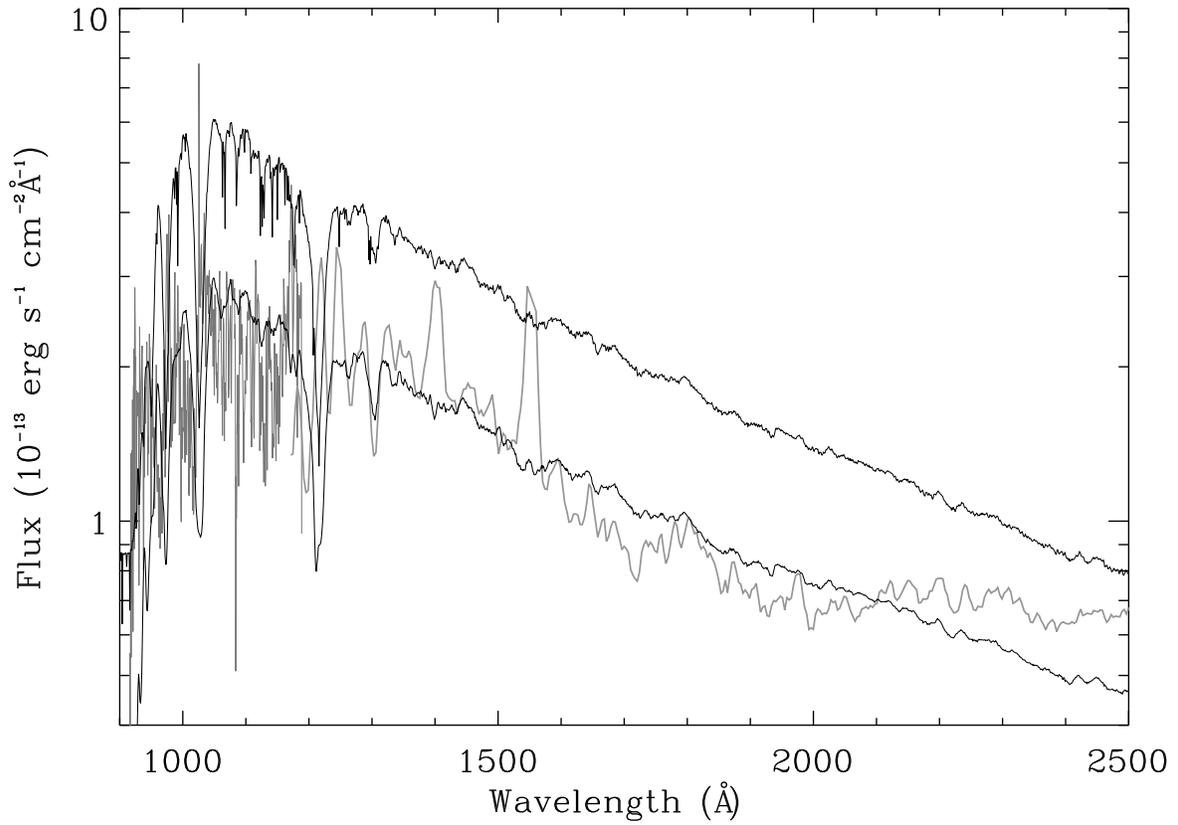}
\epsscale{1.00}
\vspace{0.5cm}
\figcaption{Grey line, ${\lambda}>1200$\AA: $FOS$ data set Y2AH0201T.
Grey line, ${\lambda}<1200$\AA: $FUSE$ orbit04 spectrum. Upper synthetic spectrum:
Model with 40,000K WD. Lower synthetic spectrum: Model with 20,000K WD.
Both synthetic spectra have been divided by $7.5{\times}10^{41}$,
corresponding to a distance of 281pc. See the text for details.
\label{F8}}
\end{figure}
%%%%%%%%%%%%%%%%%%%%%%%%%%%%%%%%%%%%%%%%%%%%%%%%%%%%%%%%%%%%%%%%%%%

%%%%%%%%%%%%%%%%%%%%%%%%%%%%%%%%%%%%%%%%%%%%%%%%%%%%%%%%%%%%%%%%%%%
\begin{figure}[tb]
\figurenum{9}
\epsscale{0.97}
\plotone{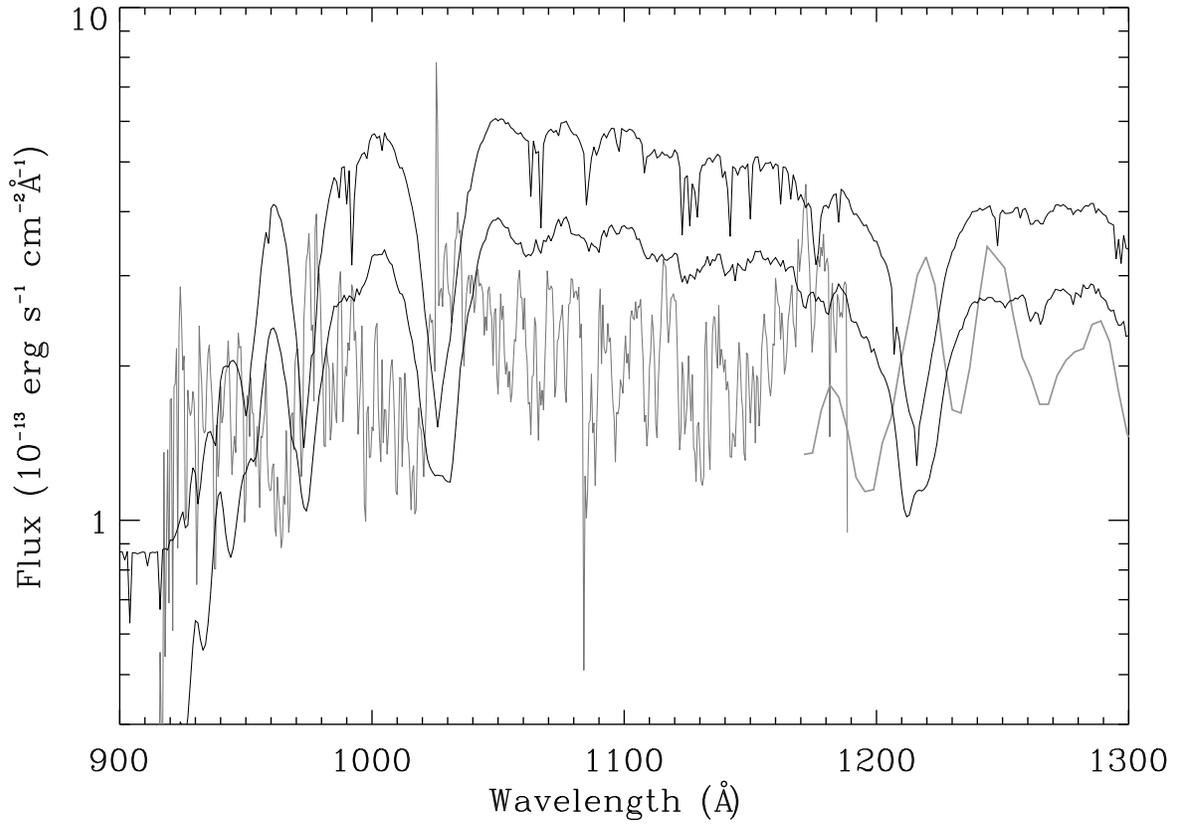}
\epsscale{1.00}
\vspace{0.5cm}
\figcaption{FUV plot of same data as Figure~8.
Lower synthetic spectrum: Model with 20,000K WD. 
Upper synthetic spectrum: Model with 40,000K WD.
See the text for details.
\label{F9}}
\end{figure}
%%%%%%%%%%%%%%%%%%%%%%%%%%%%%%%%%%%%%%%%%%%%%%%%%%%%%%%%%%%%%%%%%%%

%%%%%%%%%%%%%%%%%%%%%%%%%%%%%%%%%%%%%%%%%%%%%%%%%%%%%%%%%%%%%%%%%%%
\begin{figure}[tb]
\figurenum{10}
\epsscale{0.97}
\plotone{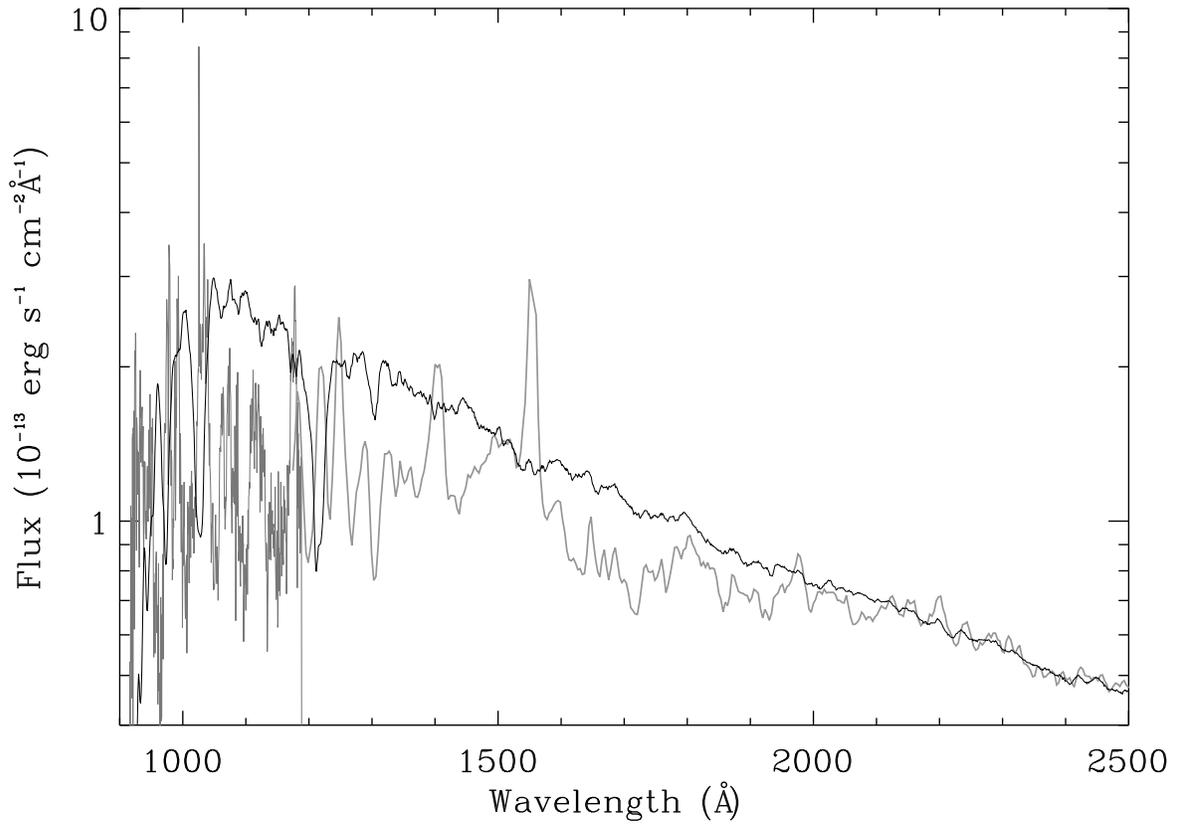}
\epsscale{1.00}
\vspace{0.5cm}
\figcaption{Grey line, ${\lambda}>1200$\AA: $FOS$ data set Y2AH0401T.
Grey line, ${\lambda}<1200$\AA: $FUSE$ orbit03 spectrum. 
The 
synthetic spectrum, using
a 20,000K WD, has been divided by $7.5{\times}10^{41}$. 
See the text for details.
\label{F10}}
\end{figure}
%%%%%%%%%%%%%%%%%%%%%%%%%%%%%%%%%%%%%%%%%%%%%%%%%%%%%%%%%%%%%%%%%%%

%%%%%%%%%%%%%%%%%%%%%%%%%%%%%%%%%%%%%%%%%%%%%%%%%%%%%%%%%%%%%%%%%%%
\begin{figure}[tb]
\figurenum{11}
\epsscale{0.97}
\plotone{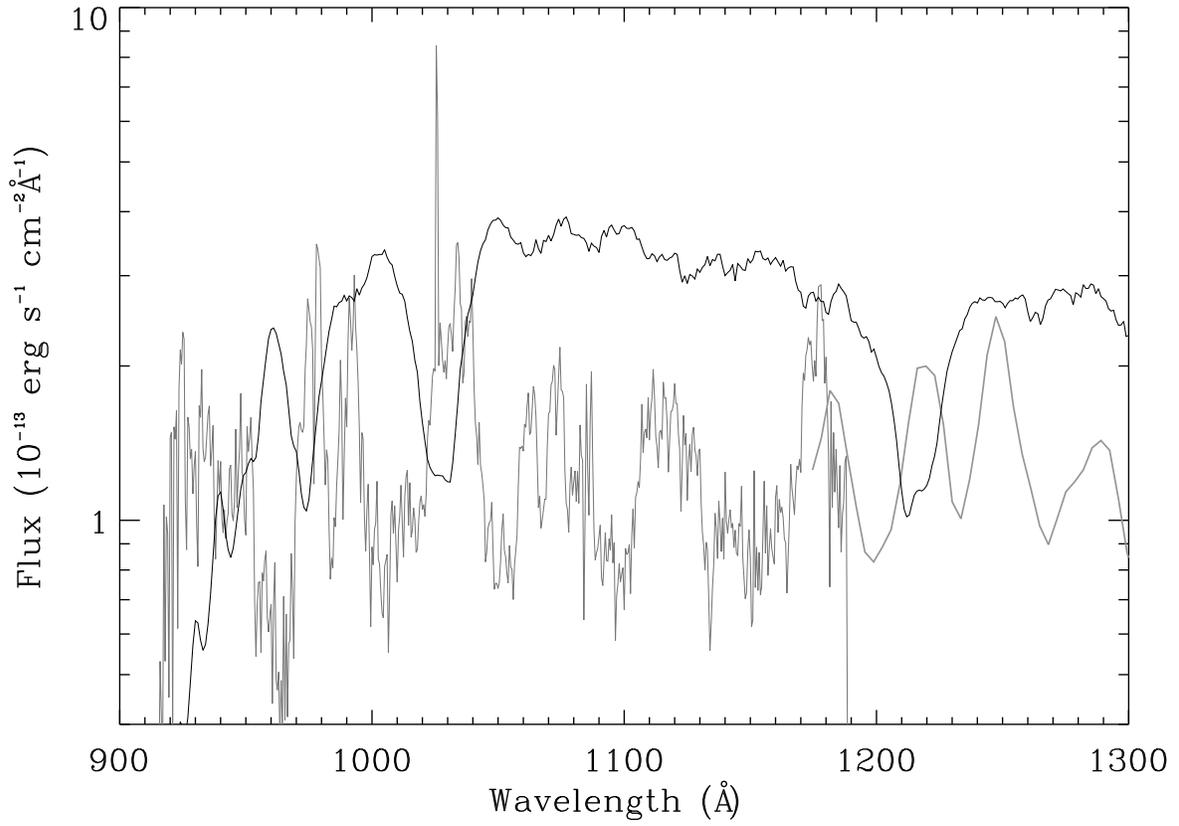}
\epsscale{1.00}
\vspace{0.5cm}
\figcaption{FUV detail of Figure~10. Note the contrast to Figure~9.
The synthetic spectrum, using
a 20,000K WD, has been divided by $7.5{\times}10^{41}$ and plotted as the
heavy line. See the text for details.
\label{F11}}
\end{figure}
%%%%%%%%%%%%%%%%%%%%%%%%%%%%%%%%%%%%%%%%%%%%%%%%%%%%%%%%%%%%%%%%%%%

%%%%%%%%%%%%%%%%%%%%%%%%%%%%%%%%%%%%%%%%%%%%%%%%%%%%%%%%%%%%%%%%%%%
\begin{figure}[tb]
\figurenum{12}
\epsscale{0.97}
\plotone{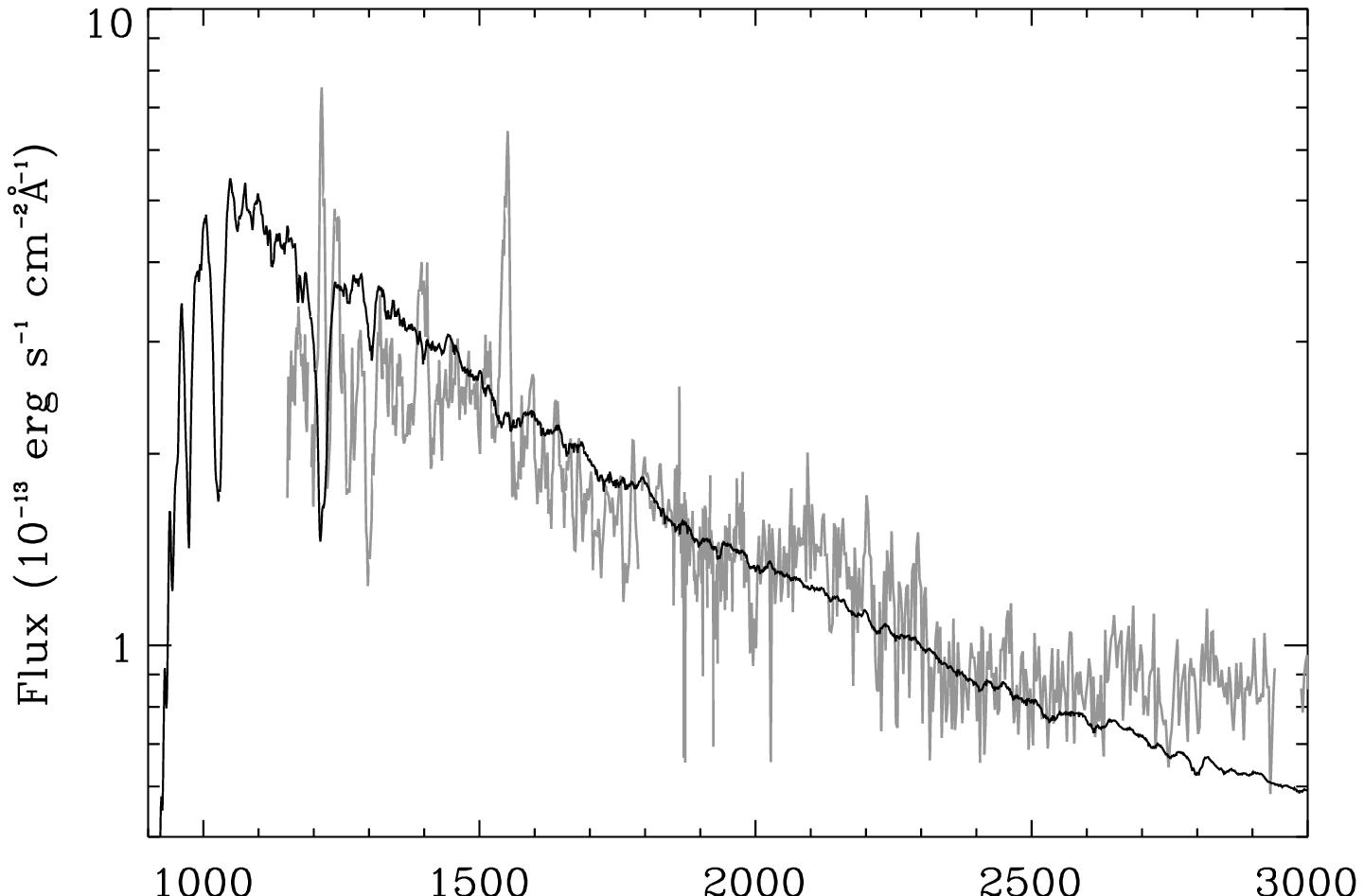}
\epsscale{1.00}
\vspace{0.5cm}
\figcaption{Grey line; SWP10371+LWR09051, observed 16/10/80.
 Heavy line: 20,000K WD and standard model accretion disk with
$\dot{M}=5.0{\times}10^{-9}M_{\odot}{\rm yr}^{-1}$ mass transfer rate
and at a distance of 234pc.	See the text for details.
\label{F12}}
\end{figure}
%%%%%%%%%%%%%%%%%%%%%%%%%%%%%%%%%%%%%%%%%%%%%%%%%%%%%%%%%%%%%%%%%%%

%%%%%%%%%%%%%%%%%%%%%%%%%%%%%%%%%%%%%%%%%%%%%%%%%%%%%%%%%%%%%%%%%%%
\begin{figure}[tb]
\figurenum{13}
\epsscale{0.97}
\plotone{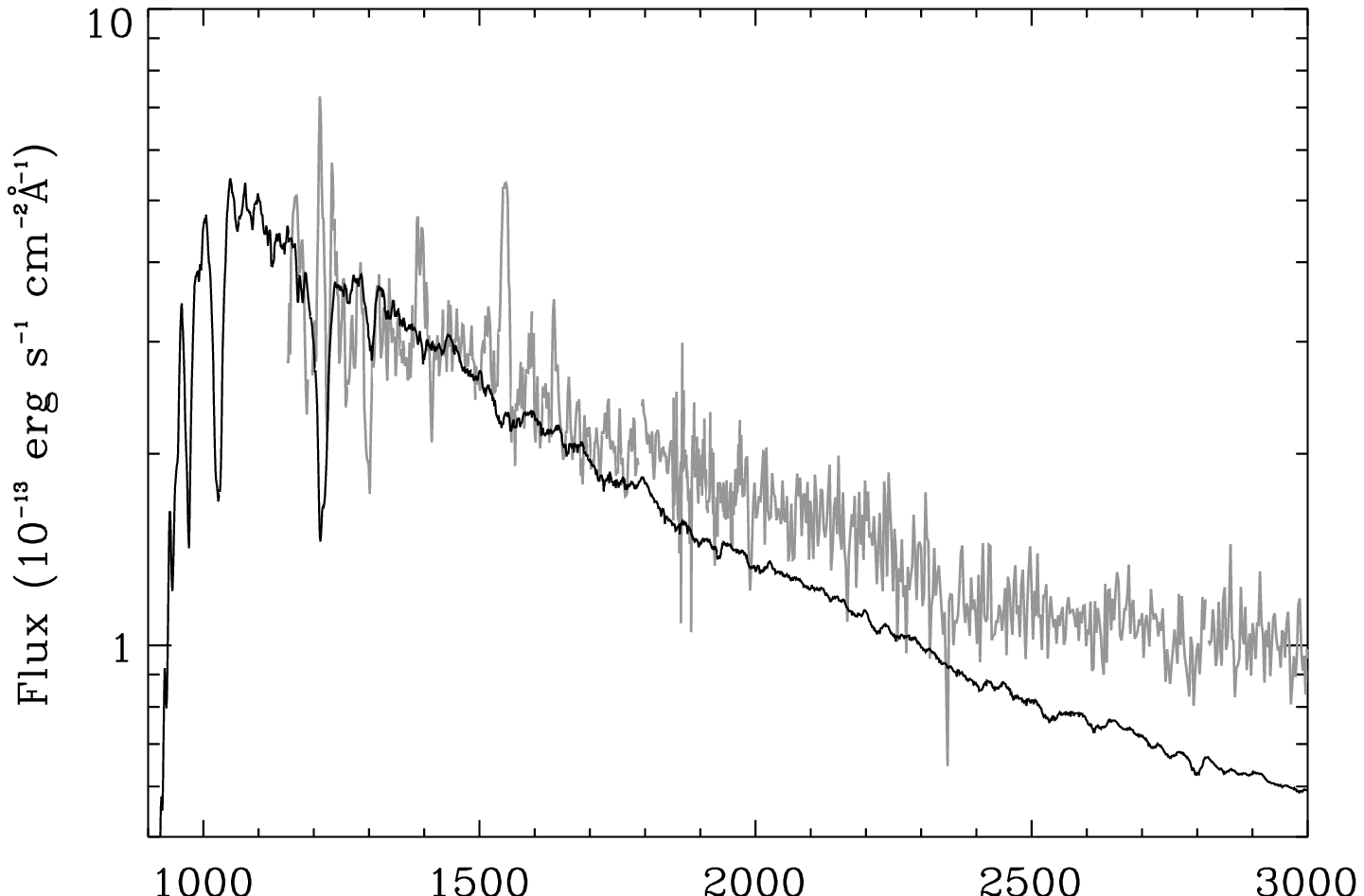}
\epsscale{1.00}
\vspace{0.5cm}
\figcaption{Grey line: SWP10128+LWR08799, observed 14/09/80. 
Heavy line: 20,000K WD and standard model accretion disk with
$\dot{M}=5.0{\times}10^{-9}M_{\odot}{\rm yr}^{-1}$ mass transfer rate
and at a distance of 234pc.	See the text for details.
\label{F13}}
\end{figure}
%%%%%%%%%%%%%%%%%%%%%%%%%%%%%%%%%%%%%%%%%%%%%%%%%%%%%%%%%%%%%%%%%%%

%%%%%%%%%%%%%%%%%%%%%%%%%%%%%%%%%%%%%%%%%%%%%%%%%%%%%%%%%%%%%%%%%%%
\begin{figure}[tb]
\figurenum{14}
\epsscale{0.97}
\plotone{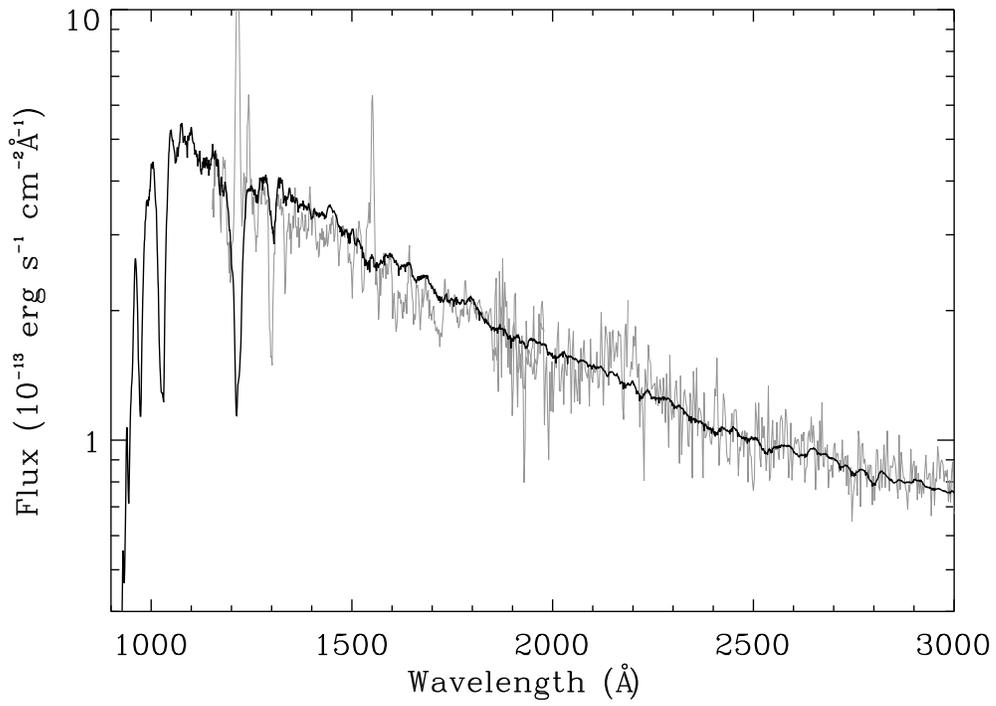}
\epsscale{1.00}
\vspace{0.5cm}
\figcaption{Grey line: SWP10677+LWR09388, observed 25/11/80. 
Synthetic spectrum: 20,000K WD and standard model accretion disk with
$\dot{M}=3.0{\times}10^{-9}M_{\odot}{\rm yr}^{-1}$ mass transfer rate
and at a distance of 165pc. 
See the text for details.
\label{F14}}
\end{figure}
%%%%%%%%%%%%%%%%%%%%%%%%%%%%%%%%%%%%%%%%%%%%%%%%%%%%%%%%%%%%%%%%%%%

%%%%%%%%%%%%%%%%%%%%%%%%%%%%%%%%%%%%%%%%%%%%%%%%%%%%%%%%%%%%%%%%%%%
\begin{figure}[tb]
\figurenum{15}
\epsscale{0.97}
\plotone{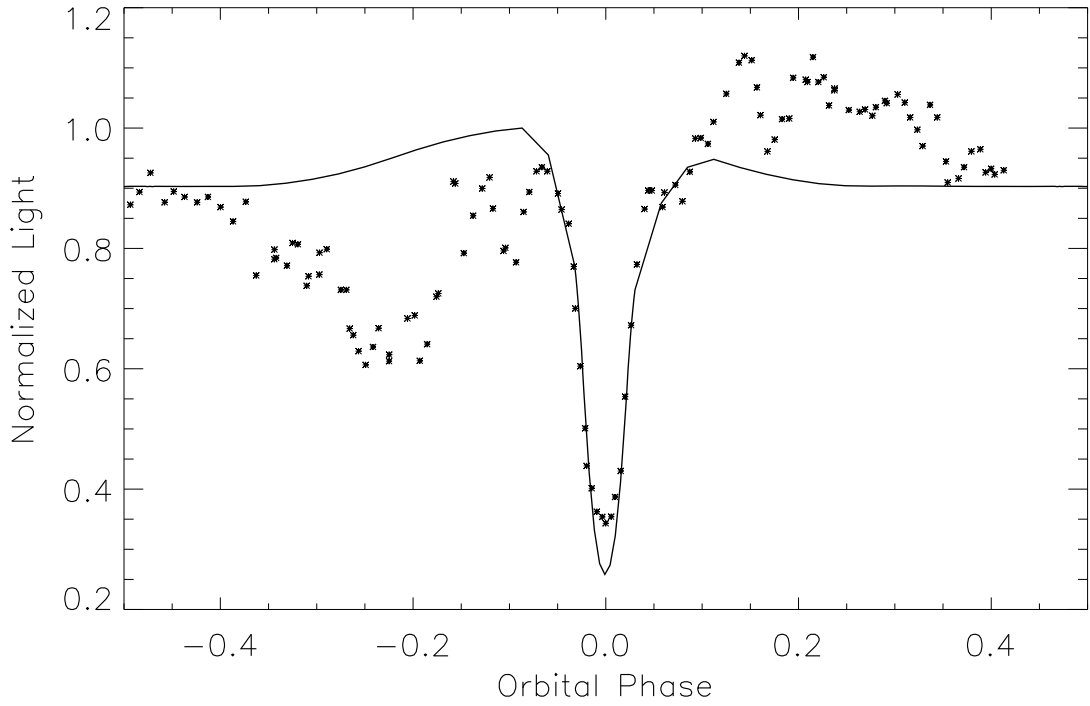}
\epsscale{1.00}
\vspace{0.5cm}
\figcaption{Fit of synthetic light curve of Figure~7 to an
observed U light curve for May 29, 1952 \citep{jph1954}. 
An extreme discrepancy from the system model, with light maximum
occurring $after$ eclipse. See the text for a proposed explanation.
\label{F15}}
\end{figure}
%%%%%%%%%%%%%%%%%%%%%%%%%%%%%%%%%%%%%%%%%%%%%%%%%%%%%%%%%%%%%%%%%%%

\end{document}